# Mapping the twist angle and unconventional Landau levels in magic angle graphene

A. Uri[1†], S. Grover[1†], Y. Cao[2†], J. A. Crosse[3], K. Bagani[1], D. Rodan-Legrain[2], Y. Myasoedov[1], K. Watanabe[4], T. Taniguchi[4], P. Moon[3,5,6], M. Koshino[7], P. Jarillo-Herrero[2]*, and E. Zeldov[1]*

[1]Department of Condensed Matter Physics, Weizmann Institute of Science, Rehovot 7610001, Israel

[2]Department of Physics, Massachusetts Institute of Technology, Cambridge, Massachusetts 02139, USA

[3]New York University Shanghai and NYU-ECNU Institute of Physics at NYU Shanghai, Shanghai, China

[4]National Institute for Material Science, 1-1 Namiki, Tsukuba, 305-0044, Japan

[5]Department of Physics, New York University, New York 10003, USA

[6]State Key Laboratory of Precision Spectroscopy, East China Normal University, Shanghai 200062, China

[7]Department of Physics, Osaka University, Toyonaka, 560-0043, Japan

†equal contribution

*corresponding authors

**Abstract**

The emergence of flat electronic bands and of the recently discovered strongly correlated and superconducting phases in twisted bilayer graphene crucially depends on the interlayer twist angle upon approaching the magic angle $\theta_M \approx 1.1°$. Although advanced fabrication methods allow alignment of graphene layers with global twist angle control of about 0.1°, little information is currently available on the distribution of the local twist angles in actual magic angle twisted bilayer graphene (MATBG) transport devices. Here we map the local $\theta$ variations in hBN encapsulated devices with relative precision better than 0.002° and spatial resolution of a few moiré periods. Utilizing a scanning nanoSQUID-on-tip, we attain tomographic imaging of the Landau levels in the quantum Hall state in MATBG, which provides a highly sensitive probe of the charge disorder and of the local band structure determined by the local $\theta$. We find a correlation between the degree of twist angle disorder and the quality of the typical MATBG transport characteristics. However, even state-of-the-art transport devices, exhibiting pronounced global MATBG features, such as multiple correlated insulator states, high-quality Landau fan diagrams, and superconductivity, display significant variations in the local $\theta$ with a span that can be close to 0.1°. Devices may even have substantial areas where no local MATBG behavior is detected, yet still display global MATBG characteristics in transport, highlighting the importance of percolation physics. The derived $\theta$ maps reveal substantial gradients and a network of jumps. We show that the twist angle gradients generate large unscreened electric fields that drastically change the quantum Hall state by forming edge states in the bulk of the sample, and may also significantly affect the phase diagram of correlated and superconducting states. The findings call for exploration of band structure engineering utilizing twist-angle gradients and gate-tunable built-in planar electric fields for novel correlated phenomena and applications.

Strong electronic correlations arise in twisted bilayer graphene when the low energy bands become exceedingly narrow in the vicinity of the magic angle (MA) [1–7]. The initial estimates of the bandwidth of these flat bands assumed a rigid and uniform rotation between the two graphene sheets leading to a moiré pattern [8–11]. Recent band structure calculations have shown, however, that twist angle relaxation within a single supercell (~13 nm for $\theta$~1.1°), which maximizes the energetically favorable AB Bernal stacking regions and minimizes the AA stacking, results in electronic reconstruction that significantly modifies the band structure [12,13]. Since the band structure of the flat bands is determined on a scale of several supercells, similarly to the predicted strong effects of heterostrain [14,15], twist angle gradients on such length scales should modify the single-particle band structure and induce symmetry breaking, possibly leading to properties that have not been considered so far. Moreover, since correlated phenomena may occur due to electronic interactions on distances larger than the supercell, the short-range twist angle variations may affect the stability of the competing orders, enriching the phase diagram of the correlated states.

Scanning tunneling microscopy (STM) studies of bilayer graphene have shown that the local twist angle can vary substantially between different places in the same sample and have resolved stacking faults and structural defects [16–22]. Determining $\theta$ by STM requires atomic scale resolution and therefore imaging of an entire device is not usually feasible. In addition, state-of-the-art MATBG samples require hBN encapsulation which prevents STM investigation of actual transport devices. Large inhomogeneities and extensive networks of stacking faults in bilayer graphene have also been observed by transmission electron microscopy (TEM) [13,23–25]. Although the TEM allows inspection of relatively large areas, determination of the local $\theta$ requires large supercells limiting the twist angle to below 1° [13]. Consequently, spatial maps of $\theta(r)$ in MATBG devices have not been attained so far and no relation between the global transport properties and the distribution of the local twist angles has been established.

MATBG samples encapsulated in hBN were fabricated using the tear-and-stack technique (see SI1) [26,27]. Figure 1d presents transport measurements of $R_{xx}$ in device *B* vs. backgate-controlled carrier density $n_e$ and applied perpendicular magnetic field $B_a$, showing characteristic MATBG features [1–7], including superconductivity (SI2), correlated insulator states at integer fractions of $n_s$ (four electrons per moiré supercell), and Landau fans, from which a global twist angle $\theta = 1.06°$ is derived (SI2). Locally, however, the twist angle $\theta(r)$ varies substantially. As we describe in detail below, the QH state in the presence of twist angle gradients, $\nabla\theta(r)$, consists of alternating compressible and incompressible strips in which the Fermi energy $\varepsilon_F$ resides correspondingly either within the Landau levels (LL) or in the energy gap between them, carrying ground state equilibrium currents $I^{com}$ and $I^{inc}$ (Fig. 1a and SI5). These two types of current of topological and nontopological origin respectively, were recently studied in detail in the QH state in monolayer graphene [28]. In contrast to the conventional QH, in MATBG these strips are present in the bulk of the sample, and moreover, the incompressible strips are typically very narrow (~50 nm width). We detect these narrow strips of intense $I^{inc}$ by utilizing a superconducting quantum interference device fabricated on the apex of a sharp pipette (SQUID-on-tip, SOT, Fig. 1a) [29]. The Pb SOTs, with a typical diameter $d \approx 200$ nm, are scanned at a height of $h \approx 30$ nm above the sample surface at $T = 300$ mK in $B_a \approx 1$ T (SI3). We apply a small *ac* excitation onto the *dc* backgate voltage $V_{bg}$, which causes a small *ac* displacement of the position $r_{ac}$ of the $I^{inc}$ strips (Fig. 1a). As a result, the corresponding *ac* Biot Savart magnetic field $B_z^{ac}$, measured by the SOT using a lock-in amplifier, shows sharp peaks whenever the incompressible strips pass under the tip (Fig. 1a, see SI4 for details), providing very sensitive means for nanoscale imaging of the LLs in the QH regime.

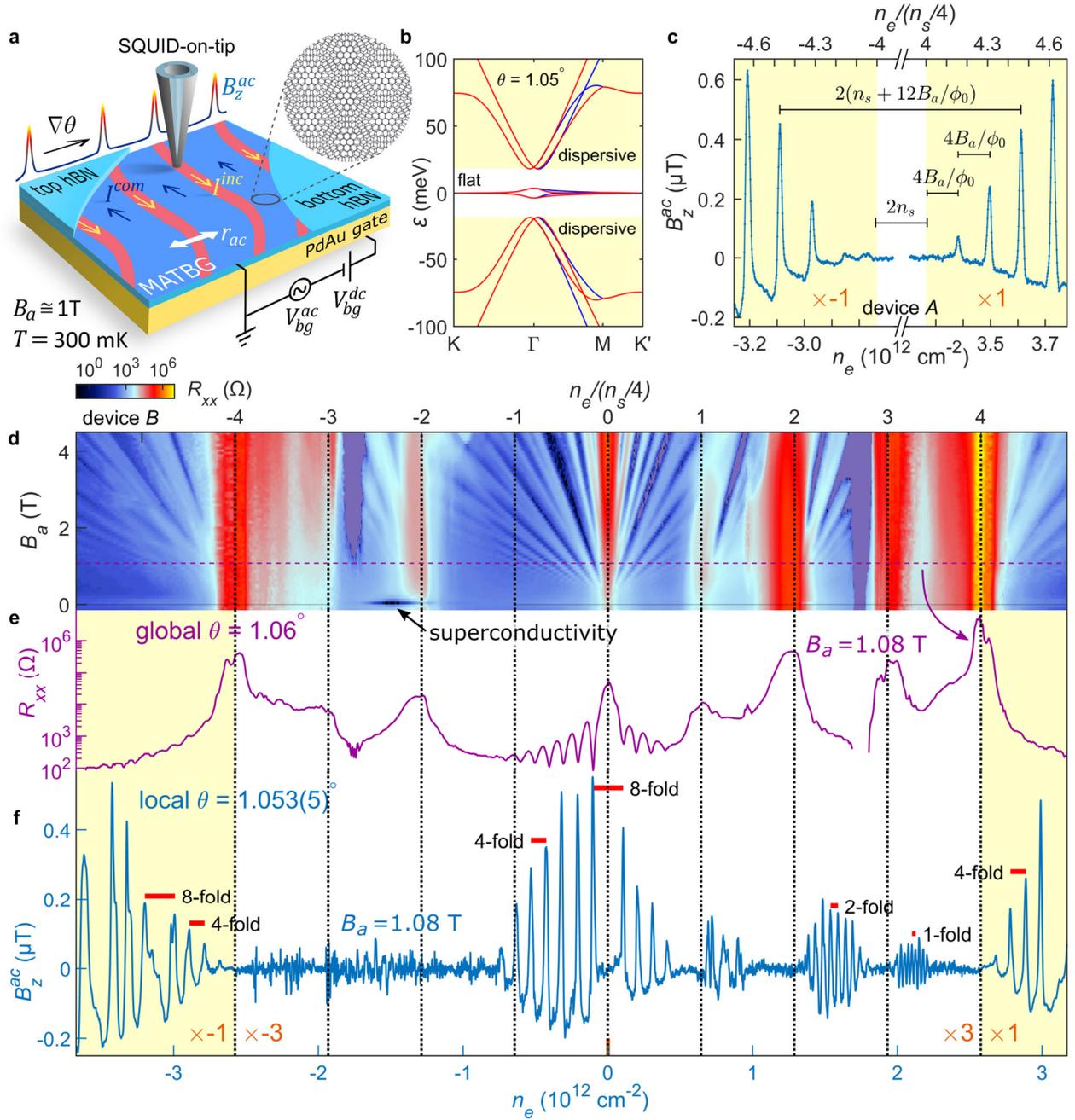

**Fig. 1. Resolving the local quantum Hall states in flat and dispersive bands in MATBG.** (**a**) Experimental setup schematics with SOT scanning over MATBG (blue) encapsulated in hBN (light blue). Voltage $V_{bg}^{dc} + V_{bg}^{ac}$ is applied between the PdAu backgate and the grounded MATBG. Twist angle gradients $\nabla\theta$ induce internal electric field and counterpropagating equilibrium QH currents $I^{inc}$ and $I^{com}$ in incompressible (red) and compressible (blue) strips respectively, flowing along equi-$\theta$ contours and measured by $B_z^{ac}$. (**b**) Calculated band structure with flat and dispersive bands. Blue and red represent the two valleys. (**c**) Zoomed-in $B_z^{ac}$ peaks in the dispersive bands for device A at $B_a = 1.19$ T, illustrating the procedure for determining the local $n_s$ and the corresponding local $\theta$ ($p$-band data multiplied by minus sign for clarity). (**d**) Global $R_{xx}$ vs. electron density $n_e$ and $B_a$ of device B showing insulating states at integer fillings, Landau fans and superconductivity. (**e**) $R_{xx}(n_e)$ at $B_a = 1.08$ T (dashed purple in (d)). (**f**) $B_z^{ac}$ measured at a point in the bulk of device B vs. $n_e$ at $B_a = 1.08$ T. The sharp $B_z^{ac}$ peaks reflect $I^{inc}$ in incompressible strips with sign determined by $\sigma_{yx}$, magnitude by LL energy gap, and separation by LL degeneracy (red bars). The dispersive bands are shaded in yellow, the signal in the flat bands is amplified 3 times, and the $p$-doped signal is multiplied by minus sign.

Figure 1f shows a sequence of such $B_z^{ac}$ peaks vs. $n_e$ for device *B*, acquired at a fixed SOT position at $B_a = 1.08$ T, along with a corresponding trace of $R_{xx}$ in Fig. 1e (see SI7 for similar data of device *A*). The positions and the magnitudes of these peaks provide wealth of information. An incompressible QH strip appears at location $r$ in the sample when the local carrier density $n_e(r)$ matches an integer number $N$ of full LLs, $|n_e(r)| = gN|B_a|/\phi_0$, where $g$ is the LL degeneracy. Hence the spacing, $\Delta n_e$, between adjacent peaks reveals the degeneracy $g$ of the LLs. The height of $B_z^{ac}$ peaks is proportional to $I^{inc} = \sigma_{yx}\Delta\varepsilon_n/e$ ($\sigma_{yx} = \nu e^2/h$ is QH conductance, $e$ – elementary charge, $\nu$ – integer filling factor, and $h$ is Planck's constant), and thus reflects the energy gap between the adjacent LLs, $\Delta\varepsilon_n = \varepsilon_{|n|+1} - \varepsilon_{|n|}$ (see SI5).

We start by inspecting high dopings, $|n_e| > n_s$, for which the Fermi level $\varepsilon_F$ resides in the dispersive bands (yellow in Fig. 1b). Figure 1c presents a zoom-in on the four lowest LLs in the electron-like (*n*) and hole-like (*p*) dispersive bands for device *A* at $B_a = 1.19$ T. The spacing between neighboring peaks is $\Delta n_e = 1.15\times 10^{11}$ cm$^{-2}$ which equals $4B_a/\phi_0$, showing that these LLs are fourfold degenerate ($\phi_0 = h/e$). The spacing between the corresponding *p* and *n* LLs equals $2(n_s(r) + 4N|B_a|/\phi_0)$ as illustrated in Fig. 1c. Since the $I^{inc}$ peaks are very sharp, measurement of this spacing renders high-accuracy determination of the local $n_s(r)$ and thus of the local twist angle $\theta(r) = a\sqrt{\sqrt{3}n_s(r)/8}$, where $a = 0.246$ nm is the graphene lattice constant. As described in SI6, this method allows determination of $\theta(r)$ with absolute accuracy of $\pm 0.005°$ and relative accuracy between different locations of $\pm 0.0002°$. In 2D scanning mode described below, we attain sensitivity of $0.007°/$Hz$^{1/2}$ and provide $\theta(r)$ maps with relative accuracy of better than $\pm 0.001°$.

Rather than parking at a fixed location, Fig. 2a shows $B_z^{ac}$ in device *A* acquired upon scanning the SOT along the white dashed line in Fig. 3a and sweeping $V_{bg}$, revealing very rich patterns. Focusing first on the dispersive bands (yellow shading), rather than being fixed, the LLs vary strikingly in space. Moreover, the degeneracy of the higher LLs toggles between 4-fold and 8-fold as a function of position, and a pronounced asymmetry between the LL structure in the *n* and *p* dispersive bands is observed.

As in Fig. 1c, by tracing the spacing between the lower LLs we derive the local $n_s(x) = C(V_{ns}(x) - V_{-ns}(x))/2$, where $V_{ns}(x)$ and $V_{-ns}(x)$ are the backgate voltages corresponding to the local filling of the flat bands $|n_e(x)| = n_s(x)$ (dashed yellow curves in Fig. 2a), and $C$ is the backgate capacitance (SI2). The attained $n_s(x)$ (Fig. 2b) varies by about $2.4\times 10^{11}$ cm$^{-2}$ corresponding to the local variation in $\theta(x)$ of 3.9% from 1.124° to 1.169° over the 2.7 µm long path (Fig. 2c). In addition to the twist angle disorder, which shifts the *p* and *n* LLs antisymmetrically in opposite directions, we also derive the local charge disorder $n_d(x)$, which shifts all the LLs symmetrically through variation of the local charge neutrality point (CNP), $n_d(x) = CV_{CNP}(x) = C(V_{ns}(x) + V_{-ns}(x))/2$. The derived charge disorder $\delta n_d(x) = n_d(x) - \bar{n}_d$ (Fig. 2d) has standard deviation of $0.8\times 10^{10}$ cm$^{-2}$ which is substantially smaller than $n_s(x)$ variation, showing that the dominant source of disorder in this MATBG device arises from $\theta(r)$ variations, as evident in Fig. 2a by the antisymmetric bending of the dispersive *p* and *n* LLs.

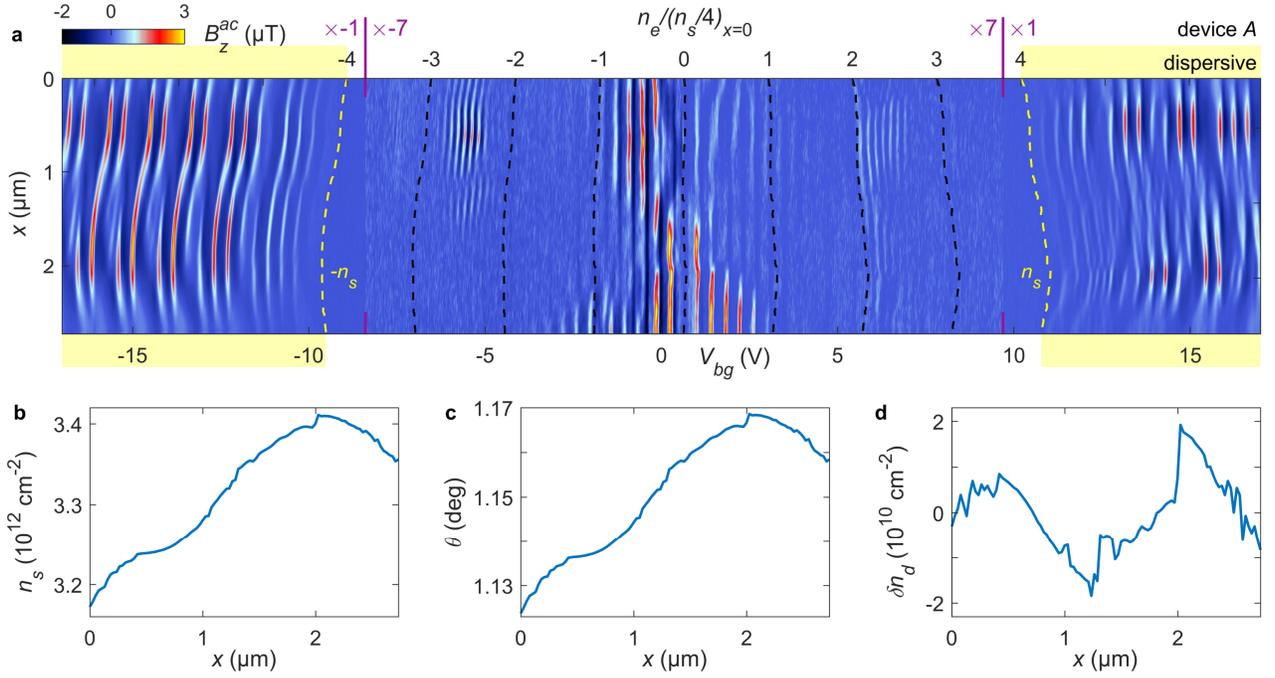

**Fig. 2. Structure of the Landau levels and derivation of the twist angle along a line scan.** (a) $B_z^{ac}(x)$ vs. $V_{bg}$ for device *A* acquired along the dashed line in Fig. 3a. The top axis denotes $n_e/(n_s(x)/4)$ for $x = 0$ and the separation between the yellow dashed lines describes the evolution of $n_s(x)$. The dispersive band regions are marked in yellow. The signal in the flat bands is amplified 7 times and multiplied by minus one for *p*-doping such that incompressible strips are bright. (**b-d**) The derived position-dependent $n_s(x)$ (**b**), $\theta(x)$ (**c**), and the charge disorder $\delta n_d(x)$ (**d**).

To derive full maps of the local twist angle $\theta(r)$ and charge disorder $\delta n_d(r)$, we acquired movies M1-M4 of $B_z^{ac}$ images upon incrementing $V_{bg}$ through the bottom of the dispersive bands (SI8). Figure 3a displays one frame from Movie M2 showing a large-area scan of device *A* (dashed rectangle in the AFM inset), while Movie M1 presents zoomed-in imaging of the central region of the device (dashed rectangle in Fig. 3a). The red stripes reveal incompressible regions carrying $I^{inc}$ while the dark blue mark the compressible areas carrying counterpropagating $I^{com}$. As $V_{bg}$ varies, the QH states move and change their shape in an intricate manner. Surprisingly, the quantum Hall edge states are present in the bulk of the sample and do not flow parallel to the sample edges as expected. Moreover, large parts of the sample do not show LLs at all. These are the regions that are either highly disordered or may have a very different twist angle, with $\theta$ either close to zero or $\theta > 1.5°$ such that the dispersive bands are reached at $V_{bg}$ outside our range. Thus, the MA physics appears only in a limited central region of the sample and does not fully extend to the edges. Figure 3e shows a larger area $B_z^{ac}$ image of device *B* displaying QH states over most of its area. Movies M3, M4 were acquired in the central part of the Hall-bar structure (dashed rectangle in the AFM inset) probing *p* and *n* dispersive bands respectively.

Using these data we generate 3D tomographic rendering of the structure of the LLs throughout the samples (SI9) that can be inspected interactively [30]. Figure 3d shows a slice of the tomographic data of device *A* (see Movie M5), revealing the layered structure of incompressible (light blue/red) and compressible (dark blue) QH regions. Strikingly, the LLs display steep slopes and numerous small jumps in the bulk of the sample, revealing that at any value of $V_{bg}$ (horizontal tomographic plane) several different LLs cross the Fermi level in the bulk of the sample never forming a well-defined single QH state. This observation explains the absence of clear conductance oscillations in the global $R_{xx}$ data in the dispersive bands in Fig. 1e in contrast to the very sharp local features.

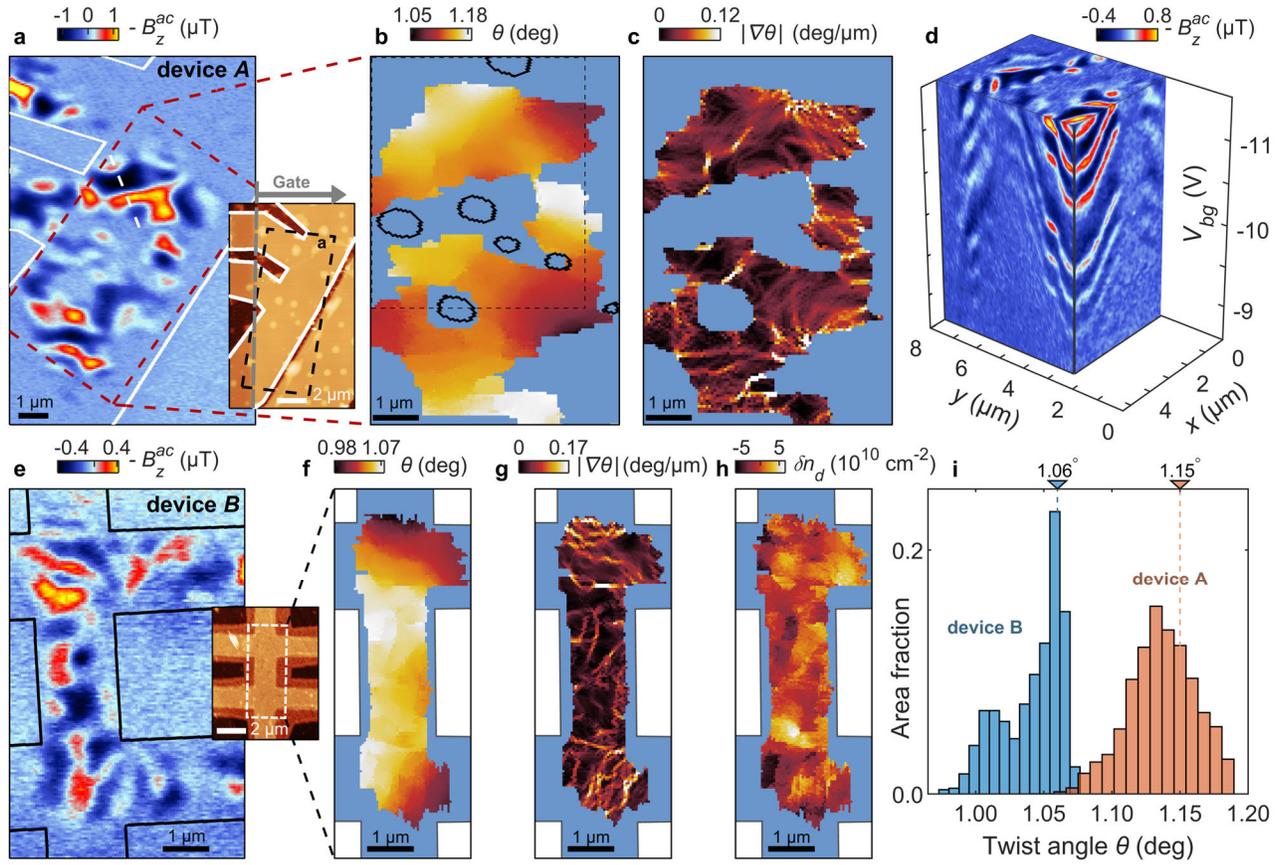

**Fig. 3. Mapping the twist angle and Landau levels in MATBG. (a)** $B_z^{ac}$ image of the dashed area in the AFM inset at $V_{bg} = -16.4$ V. Incompressible (compressible) QH regions are visible as bright blue to yellow (dark blue). Inset: AFM image of hBN encapsulated MATBG device *A* with edges outlined in white, light brown area indicating the underlying PdAu backgate, and bright spots showing bubbles. **(b)** 2D map of the twist angle $\theta(r)$ derived from tomography of Movie M1 in the dashed region in (a). Grey-blue indicates regions which do not display MATBG physics due to disorder (bubbles outlined in black) or due to a very different twist angle. The dashed area is presented tomographically in (d). **(c)** 2D map of $|\nabla\theta(r)|$ showing patches of slowly varying $\theta(r)$ and a network of abrupt $\theta$ jumps. **(d)** Slice from the tomography of device *A* showing disordered LLs in the bulk of the sample in the *p* dispersive band (see Movie M5 and [30] for interactive interface). The x-axis is flipped for clarity. **(e-g)** Same as (a-c) for device *B* with (e) acquired at $V_{bg} = -15$ V and (f) derived from tomography of Movies M3 and M4. **(h)** Charge disorder map $\delta n_d(r)$ of device *B*. **(i)** Histogram of local $\theta$ in devices *A* and *B* with dashed lines marking the global $\theta$ derived from transport.

Applying the procedure of Fig. 1c to the tomographic LL data, we derive 2D maps of the twist angle $\theta(r)$ in devices *A* and *B* (Figs. 3b,f). The grey-blue color in Fig. 3b reflects areas where no QH states were detected within the measured span of $V_{bg}$. These regions correlate with the locations of bubbles (black outlines) as revealed by the AFM image of device *A* (Fig. 3a inset). The magic angle physics is apparently absent within the bubbles as well as in their surrounding areas up to 0.5 μm from the bubble edges. The LLs are absent also in additional regions where no particular features were observed in the AFM. The map in Fig. 3b also shows that the MA regions in device *A* do not create a percolation path between the contacts. This is consistent with our transport measurements that do not show fully developed superconductivity, although correlated insulating states are present in this device (SI2). In device *B*, in contrast, four-probe transport measurements showed high quality correlated insulator states at multiple integer filling factors of the superlattice, and a zero resistance superconducting state (SI2) consistent with the observation that the MA area extends over the entire length of the central part of the device (Fig. 3f).

Figure 3h presents the derived charge disorder map $\delta n_d(r)$ in device B. The histogram in SI10 shows a standard deviation of disorder $\Delta n_d = 2.59 \times 10^{10}$ cm$^{-2}$, which is comparable to high-quality hBN encapsulated monolayer graphene devices [31] and significantly lower than in graphene on SiO$_2$ [32]. Note that in contrast to hBN encapsulated graphene, MATBG fabrication process is currently incompatible with thermal annealing procedures for disorder reduction. We observe that the charge disorder in device B is notably larger than in the MA regions in device A (Fig. 2d and SI10), which we ascribe to the fact that in contrast to the latter, device B did not undergo surface residues cleaning by AFM.

The MA regions show significant variations in the twist angle (Fig. 3i histogram). The $\theta(r)$ spans a range of 0.13° (1.05° to 1.18° with standard deviation 0.025°) in device A (Fig. 3b) and 0.10° in device B (0.98° to 1.08°, standard deviation 0.022°, Fig. 3f). Moreover, the topography of $\theta(r)$ is nontrivial with numerous peaks and valleys, as well as saddle points. Since the LLs in the dispersive band follow the bottom of the band, $n_s = 8\theta^2/(\sqrt{3}a^2)$, the LLs first appear at the minimum of $\theta(r)$ landscape, which for device A occurs in the lower-right corner (dark brown in Fig. 3b). This behavior is clearly visible in Movie M1 where arc-like incompressible strips (bright) first appear at this corner and upon increasing $|V_{bg}|$ "climb" the amphitheater-like $\theta(r)$ landscape following the equi-$\theta(r)$ contours. Similar behavior is observed in other regions with particularly interesting dynamics occurring at the saddle points as described in SI8.

The $\theta(r)$ derived in Figs. 3b,f appears to be rather smooth with typical gradients of ~0.05°/μm. Remarkably, Figs. 3c,g presenting the gradient, $|\nabla\theta(r)|$, reveal that variations in $\theta(r)$ partially occur through a network of small steps of variable sizes reaching 0.01°. The derived pattern strongly resembles the stacking fault networks in bilayer graphene observed by TEM [13,23–25]. These steps cause the stepwise jumps in the LLs visible in the tomographic view in Figs. 3d and S8a,c. This finding implies that the smooth variations in $\theta(r)$ are accompanied by occasional small abrupt changes across stacking faults that relax the tensile and shear stress in MATBG devices.

The revealed variations and gradients in $\theta(r)$ may have significant implications on the phase diagram and transport properties of MATBG. Connecting regions of different $\theta$ (Fig. 4a) is akin to attaching materials with different band structures and work functions (Fig. 4b), resulting in charge transfer and creation of internal electric fields (Figs. 4c-e). In the absence of charged impurities, at $V_{CNP}$ the sample is neutral everywhere and there are no in-plane electric fields, $E_\parallel(r) = 0$. Upon doping the sample with average $\bar{n}_e \cong CV_{bg}$, locations with different $\theta(r)$ and hence different density of states (DOS) will translate this $\bar{n}_e$ into different chemical potential $\mu(r)$. Since at thermal equilibrium the Fermi level has to be uniform, $\varepsilon_F = \mu + qV = 0$, (the last equality reflects the grounding of the device) variation in $\mu(r)$ imposes variation in electric potential $V(r) = -\mu(r)/q$, which generates in-plane electric field $E_\parallel = -\nabla V$ that cannot be screened (here $q = \pm e$ is the carrier charge). Utilizing the DOS derived from band structure calculations (SI11), Figs. 4c-e present a self-consistent numerical calculation (SI12) of $V(x)$, $E(x)$, and $\delta n_e(x) = n_e(x) - \bar{n}_e$ at $B_a = 0$ for the case of linearly varying $\theta(x)$ with $\nabla\theta = 0.025°/\mu m$ comparable to the measured average gradients in Figs. 3c,g. A significant electric field $E \cong 0.4$ kV/m is formed in the region of varying $\theta(x)$, while the accompanying charge redistribution remains very small, $\delta n_e/\bar{n}_e \cong 3\cdot 10^{-5}$.

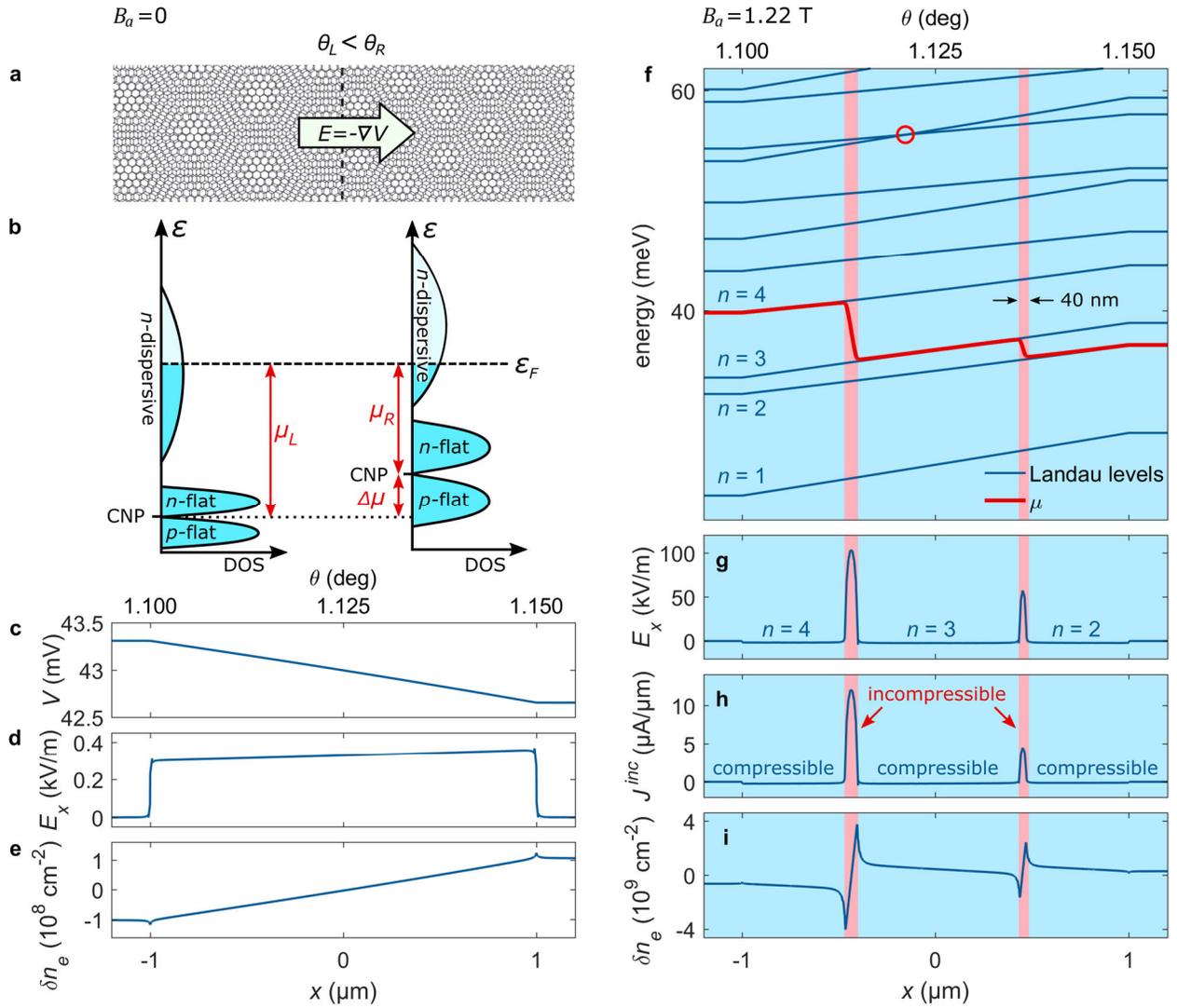

**Fig. 4. Unconventional QH state, internal electric fields and persistent currents induced by twist-angle gradients.** (**a**) Cartoon of a MATBG sample with two regions of different twist angles. (**b**) Schematic DOS of the two connected regions with different $n_s$ showing the difference in chemical potentials $\mu_L$ and $\mu_R$. (**c-e**) Finite element calculation of the potential $V(x)$ (**c**), electric field $E_x(x)$ (**d**), and carrier concentration $\delta n_e = n_e(x) - \bar{n}_e$ (**e**) for the case of linear change in $\theta$ from 1.10° to 1.15° with 0.025°/μm gradient and average carrier density $\bar{n}_e$= 3.25×10$^{12}$ cm$^{-2}$ at $B_a = 0$. (**f-i**) Single particle calculation of LL energies $\varepsilon_n(x)$ in the $n$ dispersive band and of self-consistent chemical potential $\mu(x)$ (**f**), $E_x(x)$ (**g**), $J^{inc}(x)$ (**h**), and $\delta n_e(x)$ (**i**) at $B_a = 1.22$ T. Sharp peaks in $E_x$ and $J^{inc}$ correspond to narrow (~50 nm wide) incompressible strips (shaded red) that are observed experimentally as peaks in $B_z^{ac}$.

In presence of magnetic field, the DOS variations induced by $\theta(r)$ give rise to gradients in the LL energies $\varepsilon_n$ as depicted by the blue lines in Fig. 4f (SI12). As a result, a highly unusual QH state emerges in which instead of being restricted to the edges, the QH edge states are formed in the bulk creating interlaced compressible and incompressible strips with different integer filling factors (Figs. 4g-i). In contrast to the conventional QH, in which the edge states must form closed loops, here they seem to terminate in the bulk upon reaching non-MA regions (Movies M1-4). Moreover, instead of the usually required constant carrier density in the incompressible regions, in presence of a $\theta$ gradient the density varies following the variation in $n_s(r)$ (Fig. 4i). This gradient also causes accidental LL crossings (red circle in Fig. 4f) giving rise to occasional eight-fold degenerate LLs in the dispersive bands as observed in Figs. 1f, 2a and S7. These findings provide an important insight into the fact that transport measurements of MATBG commonly show SdH oscillations without displaying full conductance quantization even at high magnetic fields [1–7]. Edge states in the bulk should disappear at high enough field as the LL degeneracy $4B_a/\phi_0$ exceeds the $n_s(r)$ variations, retaining QH quantization. Figure 4g also shows large electric fields (~$10^5$ V/m) formed in the incompressible strips giving rise to very narrow channels of persistent current $I^{inc}$ (Fig. 4h) that we image as $B_z^{ac}$ peaks. The typical width of ~50 nm of the channels along with the local $|\nabla\theta(r)|$ determines the spatial resolution of our $\theta(r)$ mapping (SI6).

Finally, we discuss the rich structure observed in the flat bands in Figs. 1 and 2. In contrast to transport measurements that resolve SdH oscillations at high fields where some of the degeneracies may be lifted, we probe the LLs locally at relatively low fields. The 0$^{th}$ LL at CNP is apparently eight-fold degenerate followed by four-fold degenerate LLs on both sides (Fig. 1f). It has been argued that such degeneracy indicates breaking of $C_3$ symmetry [33,34], which may in turn be triggered by the observed $\theta$ gradients. Figures 1f and 2a show that these LLs are sometimes observed to extend beyond $n_s/4$ on both *p* and *n* sides, while at other locations new irregular LLs seem to emerge for *n* doping above $n_s/4$, as visible in Fig. 1f. The LLs clearly reappear above $n_s/2$ for both dopings, showing degeneracy of 2 (Fig 2a and S7). We occasionally observe single-fold LLs above $3n_s/4$ for both dopings as seen in Figs. 1f and S7. We also observe that the amplitudes of the $I^{inc}$ peaks, which are proportional to energy gaps $\Delta\varepsilon_n$, emanating from different integer fillings, often follow a smooth envelope. This indicates that the energy gaps between consecutive LL are of similar, rather than alternating, magnitudes, indicating full lifting of a degeneracy.

The appearance of correlated physics in devices with a wide $\theta$ span of order ~0.1° may be explained by either a tolerance of MA physics to the exact $\theta$ or by percolating paths along very specific $\theta_M$. The fact that both of our devices show global MATBG physics, while having only a small overlap in their histograms in Fig. 3i, supports the former. Figure 2a shows, however, that LLs near $n_s/4$ and $n_s/2$ are quite discontinuous and those above $3n_s/4$ appear only at a few locations, indicating the fragility of the correlated states to twist angle disorder. It should be noted that $\theta(r)$ is a new type of disorder which is different from the more common kinds due to the fact that it changes the local band structure and induces local electric fields, the effects of which grow with $|n_e|$, explaining the higher visibility of the Landau fan near CNP in transport and the enhanced sensitivity of the correlated states and dispersive bands to this disorder.

Our finding that the QH state is profoundly modified by the twist-angle gradients and by the accompanying large internal electric fields suggests that the stability and electronic properties of other correlated phases in MATBG, including superconductivity and magnetism, may also be strongly affected by the twist angle landscape. The generated gate-tunable intrinsic in-plane electric fields may also be of practical importance for photovoltaic and thermoelectric applications of atomically thin twisted van der Waals materials.

**Acknowledgments:** We thank A. Stern and E. Berg for valuable discussions and M. F. da Silva for constructing the COMSOL simulations. This work was supported by the Sagol WIS-MIT Bridge Program, by the European Research Council (ERC) under the European Union's Horizon 2020 research and innovation program (grant No 785971), by the Israel Science Foundation ISF (grant No 921/18), by the Minerva Foundation with funding from the Federal German Ministry of Education and Research, and by the Leona M. and Harry B. Helmsley Charitable Trust grant 2018PG-ISL006. Y.C., P.J.-H. and E.Z. acknowledge the support of the MISTI (MIT International Science and Technology Initiatives) MIT–Israel Seed Fund. Work at MIT was supported by the National Science Foundation (DMR-1809802), the Center for Integrated Quantum Materials under NSF grant DMR-1231319, and the Gordon and Betty Moore Foundation's EPiQS Initiative through Grant GBMF4541 to P.J.-H. for device fabrication, transport measurements, and data analysis. This work was performed in part at the Harvard University Center for Nanoscale Systems (CNS), a member of the National Nanotechnology Coordinated Infrastructure Network (NNCI), which is supported by the National Science Foundation under NSF ECCS award no. 1541959. D.R.-L acknowledges partial support from Fundaciò Bancaria "la Caixa" (LCF/BQ/AN15/10380011) and from the US Army Research Office grant no. W911NF-17-S-0001. M.K. acknowledges the financial support of JSPS KAKENHI Grant No. JP17K05496. J.A.C and P.M. were supported by NYU Shanghai (Start-Up Funds), NYU-ECNU Institute of Physics at NYU Shanghai, and New York University Global Seed Grants for Collaborative Research. J.A.C. acknowledges support from National Science Foundation of China Grant No. 11750110420. This research was carried out on the High Performance Computing resources at NYU Shanghai. K.W. and T.T. acknowledge support from the Elemental Strategy Initiative conducted by the MEXT, Japan, A3 Foresight by JSPS and the CREST (JPMJCR15F3), JST.

**Author contributions:** A.U., S.G. and E.Z. designed the experiment. A.U., S.G. and Y.C. performed the measurements. A.U. and S.G. performed the analysis. Y.C., D.R.-L and P.J.-H designed and provided the samples and contributed to the analyses of the results. K.B. fabricated the SOTs. Y.M. fabricated the tuning forks. J.A.C. performed the tight binding calculations with P.M. and M.K., and K.W. and T.T. fabricated the hBN. A.U., S.G. and E.Z. wrote the manuscript. All authors participated in discussions and in writing of the manuscript.

**Data availability:** The data that support the findings of this study are available from the corresponding authors on reasonable request.

## Supplementary Information

### SI1. Device fabrication

The MATBG devices were fabricated using previously reported 'tear & stack' technique [26,27,35]. We first exfoliate monolayer graphene and hBN of 10 to 50 nm thickness on $SiO_2$/Si substrates, annealed at 350° C (for hBN only) and selected using optical microscopy and atomic force microscopy. Only flakes without wrinkles and bubbles are used. PC/PDMS polymer stack on a glass slide mounted on a micro-positioning stage is used to pick up a ~10 nm thick hBN flake. The edge of the hBN flake is then used to tear a graphene flake. The substrate is rotated by 1.1° to 1.2°, followed by pickup of the other piece of graphene. The resulting stack is encapsulated with another hBN flake of thicknesses of 30 to 70 nm which has been put onto a metallic gate made of evaporated Cr/PdAu. The device geometry is defined by electron-beam lithography and reactive ion etching, only keeping relatively clean regions. Electrical contacts to the MATBG were made by one-dimensional edge contact method [36].

Optical images of devices *A* and *B* are shown in Figs. S1a,b respectively. Device *A* was fabricated on a degenerately doped Si substrate with 300 nm $SiO_2$. The MATBG resides partly on $SiO_2$ and partly on the evaporated metallic backgate (light brown in Fig. S1a). In this work, only the metallic backgate has been used for varying the carrier concentration $n_e$ and a constant voltage $V_{bg}^{Si} = 50$ V was applied to the Si backgate for keeping the rest of the sample conductive in the transport measurements. Device *B* was fabricated on an intrinsic Si substrate with a metallic backgate extending over the full size of the device (light blue in Fig. S1b).

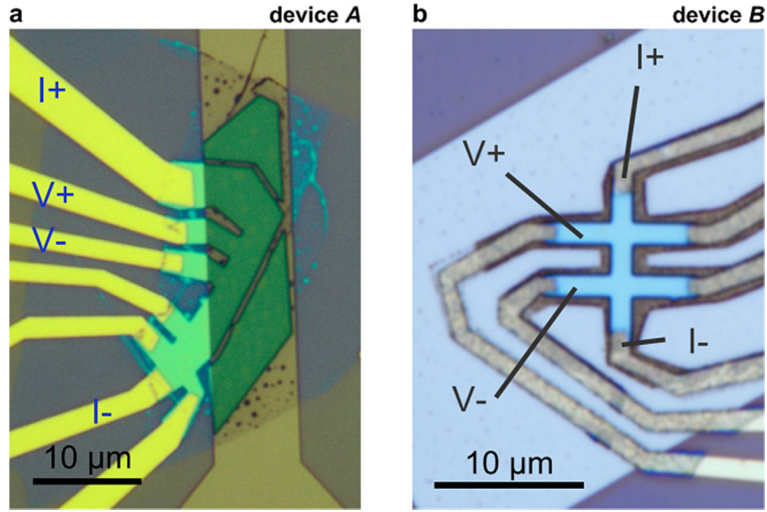

**Fig. S1. Optical image of MATBG devices.** (**a**) Optical image of device *A* showing hBN/MATBG/hBN (green), the underlying PdAu backgate (light brown), and the marked electrodes used for four-probe $R_{xx}$ measurements. (**b**) Optical image of device *B* (cyan) on the PdAu backgate (light blue) with marked electrodes.

### SI2. Transport characteristics

Four-probe resistance measurements of the samples at $T = 300$ mK are shown in Figs. S2 and S3. Both devices exhibit the common transport characteristics of correlated physics in MATBG [1–6], including $R_{xx}$ peaks at $n_s$ and its integer fractions, and Landau fans at elevated magnetic field. The slopes of the Landau fans in Figs. S2a,b were used to extract the backgate capacitances $C$ of $3.07 \times 10^{11}$ cm$^{-2}$ V$^{-1}$ (49.23 nF/cm$^2$) in device A and $2.31 \times 10^{11}$ cm$^{-2}$ V$^{-1}$ (37 nF/cm$^2$) for device B consistent with the evaluated dielectric thickness of the underlying hBN. The origins of the Landau fans were used to derive the global $n_s$ and the corresponding global $\theta = 1.15°$ for device A and $\theta = 1.06°$ in device B, in good correspondence with histograms of the local twist angle in Fig. 3i. In device A the global $\theta$ correlates with the average of $\theta(r)$ distribution, whereas in device B it is close to the upper end of the distribution function. This is consistent with the fact that the four-probe transport measurements in device B probe the central part of the Hall bar structure (Fig. S1b) where $\theta(r)$ is the highest and more uniform (Fig. 3f), while the low end tail of $\theta(r)$ distribution arises from regions that are not probed by transport.

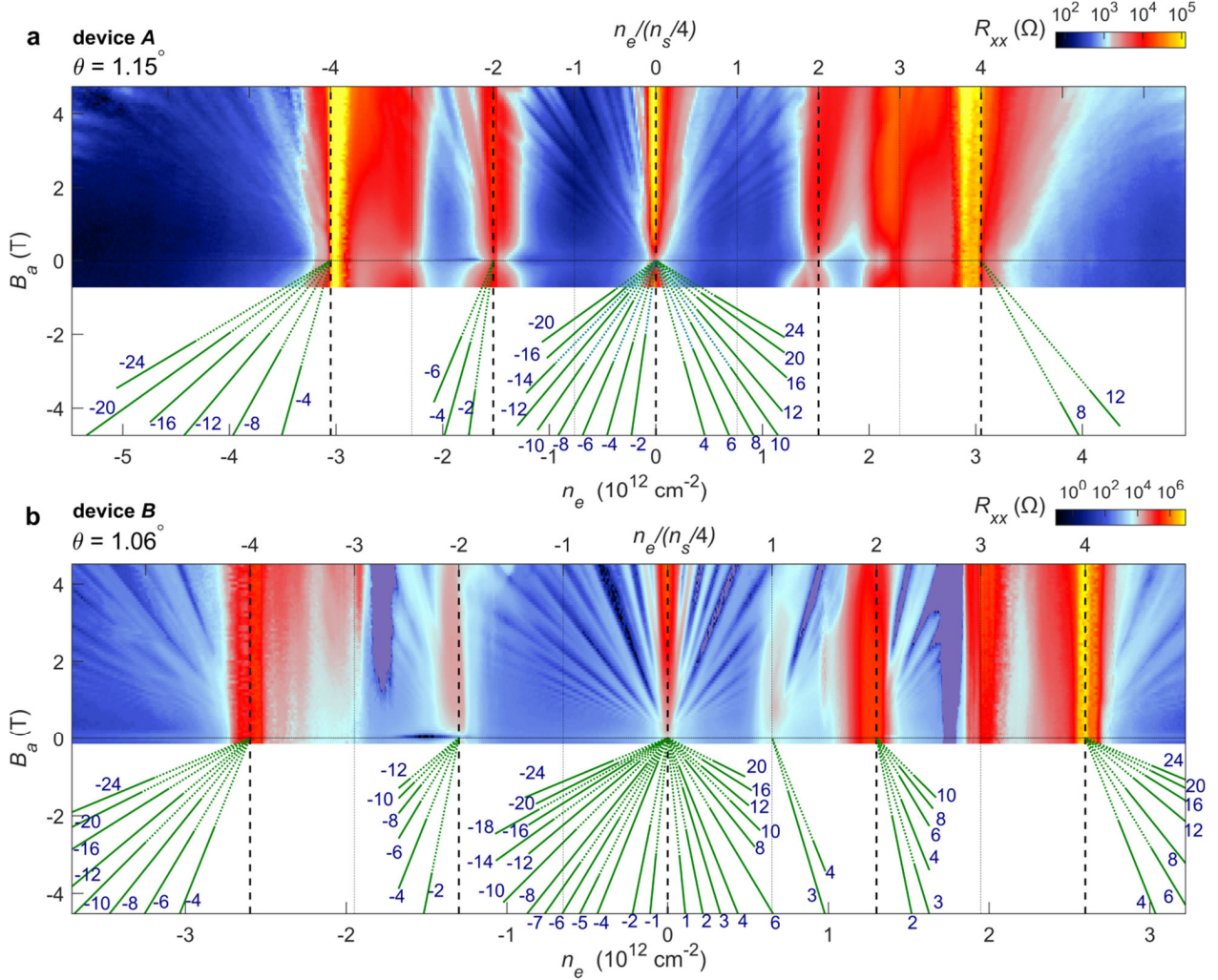

**Fig. S2. Transport measurements at $T = 300$ mK.** (a) Four-probe measurement of $R_{xx}(V_{bg})$ vs. $B_a$ in device A using excitation current of 10 nA with the corresponding traces of the Landau fan diagram at the bottom. The green solid lines show the segments that can be traced in the data and the dotted lines indicate their extrapolation to the origin. (b) Same as (a) for device B. The purple color marks the regions where the $R_{xx}$ signal was slightly negative.

In addition, in device B we observe the superconducting state in the vicinity of p-doped $n_s/2$ with zero $R_{xx}$ which becomes suppressed by small magnetic field (Fig. S3b). The critical current in the superconducting state reaches about 100 nA, as determined by the differential $dV/dI$ characteristics (Fig. S3c), and depends sensitively on the carrier density $n_e$. The observation of a fully developed superconductivity in device B is consistent with the finding of a continuous region of MA between the voltage contacts in Fig. 3f. Suppression of the resistance was also observed in device A (Fig. S3a), but the lowest $R_{xx}$ was 328 Ω, suggestive of the presence of some superconducting regions but absence of a percolation path between the voltage contacts, consistent with the $\theta$ map in Fig. 3b.

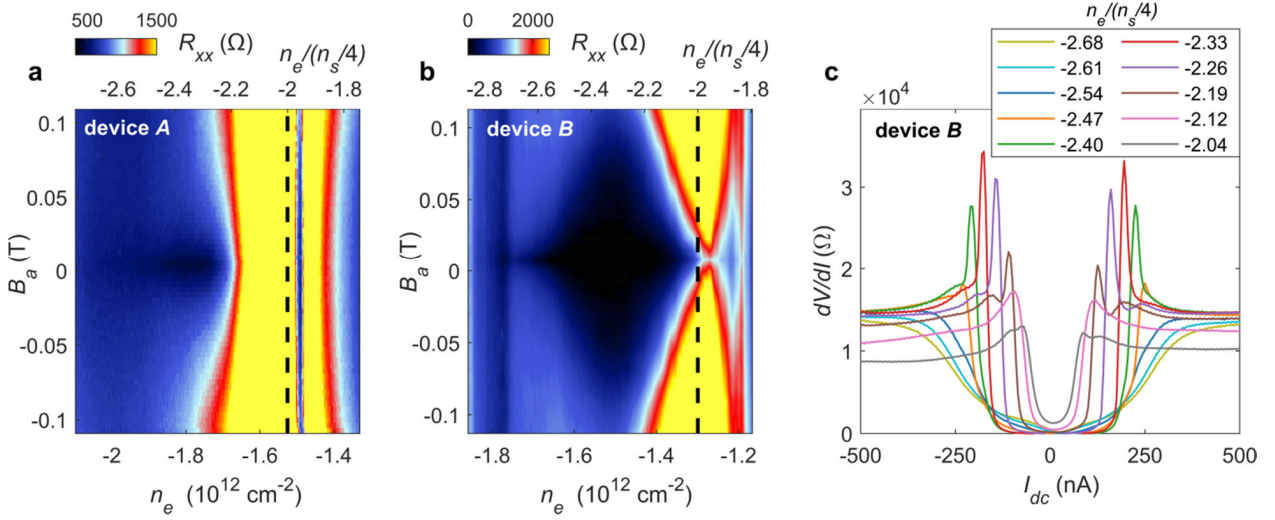

**Fig. S3. Transport measurements in the superconducting state at $T = 300$ mK.** (**a-b**) Color rendering of $R_{xx}$ measured in the vicinity of $-n_s/2$ vs. $B_a$ and $n_e$ at low fields using excitation current of 5 nA rms in device A (**a**) and 4 nA rms in device B (**b**). A zero resistance superconducting state (black) is observed in device B. (**c**) $dV/dI$ vs. $I_{dc}$ characteristics at various carrier concentrations $n_e$ in the superconducting state in device B at $B_a = 0$ T using ac excitation $I_{ac} = 10$ nA rms.

### SI3. SOT fabrication and characterization

The Pb SOTs were fabricated as described in Ref. [29] with diameters ranging from 220 to 250 nm and included an integrated shunt resistor on the tip [37]. The SOT readout was carried out using a cryogenic SQUID series array amplifier (SSAA) [38–40]. The magnetic imaging was performed in a ³He system [41] at 300 mK at which the Pb SOTs can operate in magnetic fields of up to 1.8 T. At fields $B_a \approx 1.2$ T used in this study, the SOTs displayed flux noise down to 250 n$\Phi_0$/Hz$^{1/2}$, spin noise of 10 $\mu_B$/Hz$^{1/2}$, and field noise down to 10 nT/Hz$^{1/2}$. For height control we attached the SOT to a quartz tuning fork as described in Ref. [42]. The tuning fork was electrically excited at the resonance frequency of ∼33 kHz. The current through it was amplified using a room temperature home-built trans-impedance amplifier, designed based on Ref. [43] and measured using a lock-in amplifier. The scanning was performed at a constant height of 20 to 100 nm above the top hBN surface.

### SI4. Direct current imaging technique

In order to avoid the $1/f$ noise of the SOT that is present at frequencies below ∼1 kHz, an ac signal due to backgate modulation was acquired instead of measuring the local dc $B_z(r)$. We applied a small ac excitation to the backgate (Fig. 1a), $V_{bg} = V_{bg}^{dc} + V_{bg}^{ac} \sin(2\pi f t)$, where $f \cong 3$ kHz, and the corresponding $B_z^{ac} = V_{bg}^{ac} \partial B_z/\partial V_{bg}$ was then measured by the SOT using a lock-in amplifier. Another major advantage of this modulation is that it provides a convenient method for direct imaging of the local current density $J(r)$. To demonstrate its principle, consider a $\theta$ gradient in the $\hat{x}$ direction that gives rise to a narrow strip of current of width $\Delta x$, positioned at $x_0$, and carrying a current density $J_y$ in the $\hat{y}$ direction with a total current $I_y = \Delta x J_y$ (Figs. S4a and 4h). The magnetic field $B_z(x)$ generated by the current and measured at height $h$ above it is described by the Biot Savart law (Fig. S4b). For heights $h > \Delta x$ the $B_z(x)$ is essentially governed only by the total current $I_y$ in the strip, independent of $\Delta x$. The $B_z(x)$ is an antisymmetric function with a steep slope above the current strip. Its spatial derivative $\partial B_z/\partial x$ has a sharp peak at the strip location (Fig. S4c) with a height proportional to $I_y$ and thus can provide a good means for direct imaging of the current density distribution $J_y(x)$ if the latter can be modulated in space in the $\hat{x}$ direction. The backgate voltage $V_{bg}^{ac}$ provides such spatial modulation as follows. In the presence of potential gradients, the QH edge channels flow along equipotential contours (given by equi-$\theta$ contours in absence of charge disorder). A small $V_{bg}^{ac}$ thus shifts the location of the channel by $x_0^{ac} = V_{bg}^{ac} \partial x_0/\partial V_{bg}$ in the direction parallel to the gradient and perpendicular to the current flow. So regardless of the gradient direction $\hat{x}$, the measured signal will be given by $B_z^{ac} = -x_0^{ac} \partial B_z/\partial x \propto x_0^{ac} J_y(x)$, thus providing direct imaging of the local current density. Figures S4d-f present a simulation of three counterpropagating current strips demonstrating the $B_z^{ac}$ imaging for this case.

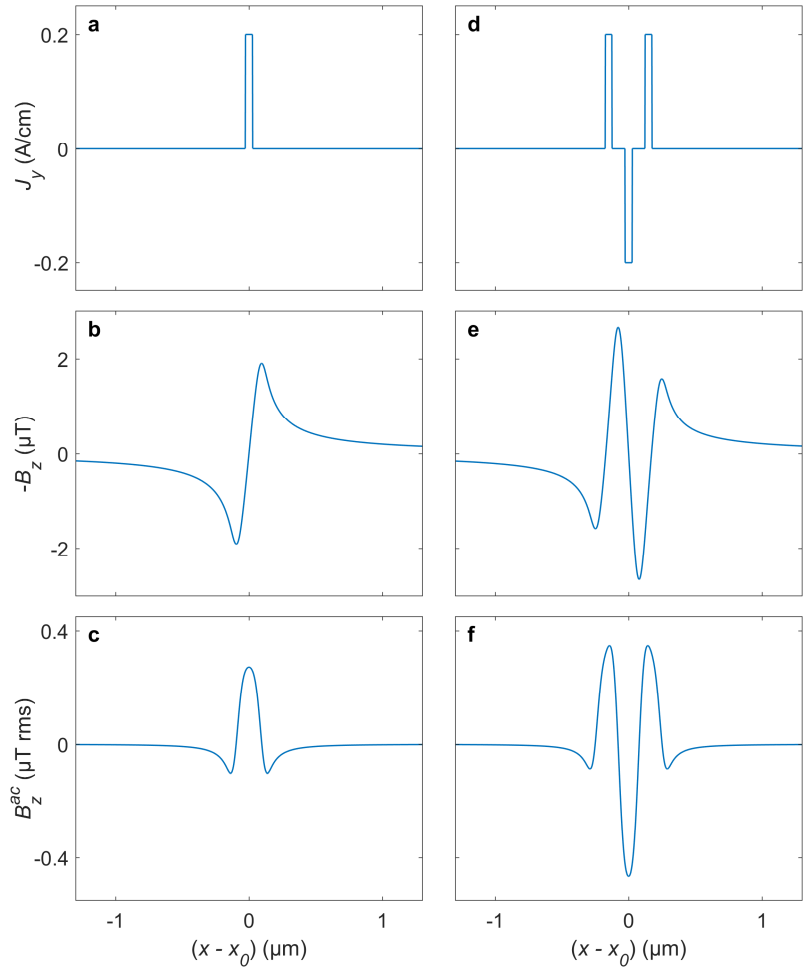

**Fig. S4**. **Numerical simulation demonstrating current imaging by measuring $B_z^{ac}$.** (**a**) Current distribution $J_y(x - x_0)$ of a 50 nm wide channel carrying $I_y = 1$ μA in the $\hat{y}$ direction. (**b**) Calculated $B_z(x - x_0)$ at a height of 30 nm above the sample convoluted with a 200 nm diameter SOT sensing area. (**c**) Calculated $B_z^{ac}(x - x_0)$ for $x_0^{ac} = 20$ nm rms spatial modulation of the channel position. (**d-f**) Same as (a-c) but for three counter-propagating currents spaced 150 nm apart.

## SI5. Currents in the compressible and incompressible QH strips

Gradients in the twist angle $\nabla\theta$ give rise to gradients in the chemical potential $\nabla\mu$ and to alternating compressible (when $\mu$ resides within a LL) and incompressible ($\mu$ in the energy gap between LLs) QH strips (Figs. 4f-i). Both regions carry current [44], however, usually only the currents in the incompressible strips, $\boldsymbol{J}^{inc} = \sigma\boldsymbol{E}$, which are of topological nature, are considered, while the nontopological currents in the compressible strips, $\boldsymbol{J}^{com} = \mu_e \nabla \times |n_e|\hat{z}$, are commonly ignored (here $\mu_e = \epsilon_k/B$ is the magnetic moment of the orbiting electron and $\epsilon_k$ is its kinetic energy [45]). These currents were recently investigated in the QH state in monolayer graphene and referred to as topological and nontopological currents [28]. The following semiclassical picture is instructive in describing $J^{inc}$ and $J^{com}$. Under strong magnetic fields and in the absence of in-plane electric fields, the charge carriers follow cyclotron orbits which can be described semiclassically as an array of circles, resulting in zero average bulk current (Fig. S5a). Applying an external in-plane electric field along the $x$-direction, $E_x = -\partial V/\partial x$, to an incompressible state, causes the circular orbitals to convert into spirals drifting along the $y$ direction, generating a current $J_y^{inc} = \sigma_{yx}E_x$ (Fig. S5b). On the other hand, applying the same external electric field to a compressible strip will result in carrier redistribution which screens the in-plane electric field. As a result the drift current vanishes, but at a cost of a non-zero gradient in the carrier density $\partial n_e/\partial x$ (Fig. S5c). Since each orbital carries a magnetic moment $\boldsymbol{\mu}_e = \mu_e \hat{z}$ which gives rise to local magnetization $\boldsymbol{m} = |n_e|\boldsymbol{\mu}_e$, the induced $\partial n_e/\partial x$ causes gradients in $\boldsymbol{m}$, and hence produces equilibrium currents through $\boldsymbol{J}^{com} = \nabla \times \boldsymbol{m}$ [44]. This accounts for a non-zero $J_y^{com} = \mu_e \partial |n_e|/\partial x$ (cyan arrows in Fig. S5c), which flows in the direction opposite to the incompressible current $J_y^{inc}$ in Fig. S5b. Since a full band does not contribute to current, $n_e$ in the above expression refers only to carriers in a partially filled band. Alternatively, $J_y^{com}$ can be understood as arising from uncompensated contributions to the current from neighboring orbitals in the presence of a gradient in the orbital density (Fig. S5c).

The total current carried by the drifting orbitals in an incompressible strip residing between two compressible regions is given by $I_y^{inc} = \int J_y^{inc} dx = \sigma_{yx}\Delta\varepsilon_n/e$, where $\Delta\varepsilon_n = \varepsilon_{|n|+1} - \varepsilon_{|n|}$ is the LL energy gap between the adjacent compressible states and $\sigma_{yx} = \nu e^2/h$ is QH conductance of the incompressible state (see Fig. 4).

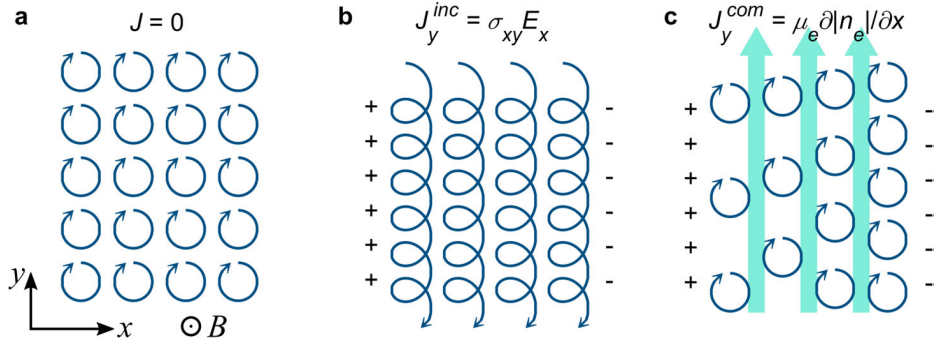

**Fig. S5. The origin of equilibrium currents in the compressible and incompressible QH states**. (**a**) Semiclassical picture of cyclotron orbits of holes with mutually canceling neighboring currents resulting in zero bulk current. (**b**) In the presence of an in-plane electric field $E_x$ (+ and − signs represent external charges) the cyclotron orbits acquire a drift velocity resulting in a non-zero $J_y^{inc}$ in the incompressible state. (**c**) In the compressible regime the external in-plane electric field is screened by establishing a charge density gradient, giving rise to $J_y^{com}$ flowing in the opposite direction (cyan arrows).

## SI6. Determination of twist angle measurement accuracy and spatial resolution

**$\theta$ accuracy.** The local twist angle is determined by the local $n_s(r)$ through $\theta(r) = a\sqrt{3n_s(r)/8}$. The incompressible $I^{inc}$ current and the corresponding peak in the $B_z^{ac}$ signal appear at specific locations where $N$ LLs in the dispersive bands are exactly fully occupied, corresponding to a density $|n_N| = C|V_{bg}^N - V_{bg}^{CNP}| = n_s + 4N|B_a|/\phi_0$ for 4-fold degenerate LLs, where $V_{bg}^N$ is the backgate voltage that corresponds to the $N^{th}$ peak. Measuring the $N^-$ and $N^+$ peaks in the $p$ and $n$ dispersive bands respectively, allows derivation of $n_s(r) = C(|V_{bg}^{+N}| + |V_{bg}^{-N}|)/2 - 4N|B_a|/\phi_0$ and therefore of $\theta(r)$. The absolute angle accuracy is thus determined by the accuracy of $C$, $B_a$, and $V_{bg}^{\pm N}$. Determination of $C$ is possible through global transport measurements and more accurately through local measurement of the spacing between any two incompressible peaks $V_{bg}^{N+1} - V_{bg}^N = g|B_a|/(\phi_0 C)$, where $g$ is the degeneracy of the Landau level considered. From this we estimate our overall absolute accuracy of determining $n_s$ to be about $\pm 1\%$, and thus absolute $\theta$ accuracy of $\delta\theta = \pm 0.005°$.

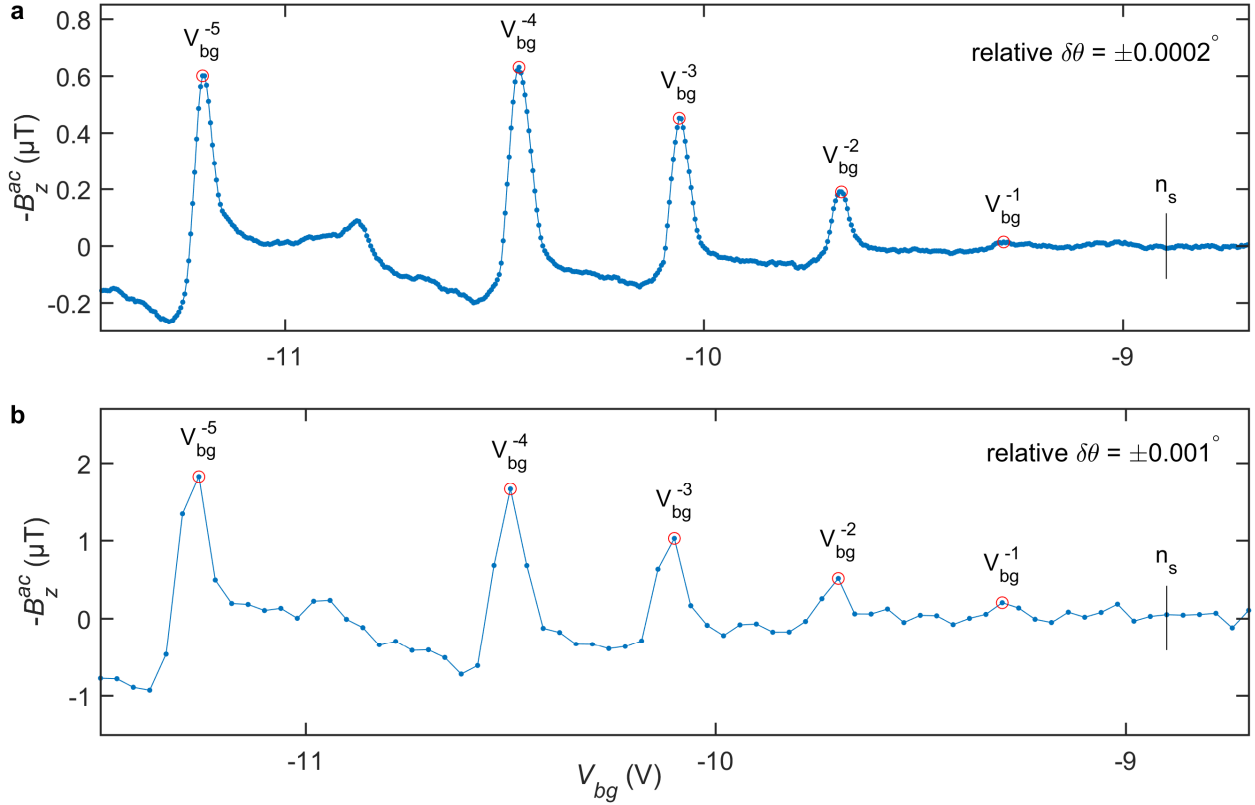

**Fig. S6. Determination of twist angle measurement accuracy.** (a) Traces of $B_z^{ac}$ vs. $V_{bg}$ in device A (from Fig. 1c) acquired with step size $\Delta V_{bg}$ of 4.7 mV and $V_{bg}^{ac}$ = 15 mV rms. The positions of the $V_{bg}^{-3}$ and $V_{bg}^{-4}$ peaks can be determined to an accuracy better than $\pm$ one step size $\Delta V_{bg}$, corresponding to relative $\theta$ accuracy of $\delta\theta = \pm 0.0002°$. (b) Same as (a) taken from Movie M1 at a pixel position $(x, y) = (2.53, 5.9)$ μm with step size $\Delta V_{bg}$ of 40 mV and $V_{bg}^{ac}$ = 35 mV rms resulting in relative $\theta$ accuracy of $\delta\theta = \pm 0.001°$ in the imaging mode. The larger $B_z^{ac}$ signal and the broader $I^{inc}$ peaks in (b) compared with (a) is due to higher $V_{bg}^{ac}$ excitation (see SI13 for measurement parameters).

In this study, however, we are particularly interested in the relative accuracy of $\theta(r)$ for comparing different locations $r$ and deriving the angle gradients $\nabla\theta$, which is determined essentially only by the measurement precision of $V_{bg}^{\pm N}$. The sharpness of the $I^{inc}$ peaks and the good signal to noise ratio of the $B_z^{ac}$ signal allow high precision measurement of $V_{bg}^{\pm N}$ as demonstrated in Fig. S6. In the stationary measurement in Fig. S6a (zoom-in of Fig. 1c), $V_{bg}$ was swept with increments $\Delta V_{bg} = 4.7$ mV demonstrating that the $V_{bg}^{-3}$ and $V_{bg}^{-4}$ peak positions can be determined to an accuracy better than $\pm$ one step size $\Delta V_{bg}$, corresponding to $\delta V_{bg}^{-4}/V_{bg}^{-4} \approx 4 \times 10^{-4}$. Since $\theta \propto \sqrt{n_s}$ we have $\delta\theta/\theta \approx 2 \times 10^{-4}$, or relative $\theta$ accuracy of $\delta\theta = \pm 0.0002°$. In movie M1, that was used to construct the full $\theta(r)$ map of device A (Fig. 3b), larger increments $\Delta V_{bg}$ of 40 mV were used (Fig. S6b), corresponding to $\theta$ accuracy $\delta\theta = \pm 0.002°$. Movies M3,4 used $\delta V_{bg} = 45$ mV constructing the $\theta(x, y)$ map of device B (Fig. 3f) with similar accuracy. Movies M3,4 contain 87 frames of 68×184 = 12,512 pixels each, that were acquired over a total of 42 hours. The $V_{bg}$ trace of each pixel therefore took $t = 12$ seconds to acquire. The $\delta\theta \lesssim \pm 0.002°$ accuracy, normalized by the pixel acquisition time provides the relative $\theta$ sensitivity per pixel in the imaging mode is better than $S_\theta^{1/2} = \sqrt{t}\delta\theta = 0.004°/\text{Hz}^{1/2}$.

**Spatial resolution of $\theta(r)$ mapping.** Our electrostatic simulations show that the typical width of the incompressible $I^{inc}$ strips is about 50 nm (Figs. 4f,h) and should be smoothened by the wavefunction width, of the order of magnetic length $l_B = \sqrt{\hbar/eB} \cong 25$ nm. Since the position $r$ of the incompressible strip provides a very accurate determination on the local $n_s(r)$ and $\theta(r)$, the width of the strip essentially determines the spatial resolution which can be smaller than the SOT diameter. The actual spatial resolution $\delta r$ is determined by the accuracy $\delta V_{bg}$ with which the $V_{bg}$ value can be assigned to the $I^{inc}$ peak at a location $r$, $\delta r = \delta V_{bg} \partial r/\partial V_{bg}$, where $\partial r/\partial V_{bg}$ is the change in position of $I^{inc}$ per change in $V_{bg}$. Since $I^{inc}$ appears at $V_{bg}(r) = (n_s(r) + 4N|B_a|/\phi_0)/C$, the space dependence enters only through $n_s(r) = n_s(\theta(r))$, therefore $\partial r/\partial V_{bg} = C(\partial\theta/\partial r)^{-1}(\partial n_s/\partial\theta)^{-1}$, where $\partial n_s/\partial\theta = 16\theta/\sqrt{3}a^2$. Using characteristic values $C = 2.5\times 10^{11}$ V$^{-1}$ cm$^{-2}$, $\delta V_{bg} = 45$ mV in the scanning mode, and $\partial\theta/\partial r = 0.057$/μm gives a resolution $\delta r = 50$ nm. Lower $\partial\theta/\partial r$ gradients result in higher $\delta r$. In such case, however, since $\theta$ varies slowly in space, a lower spatial resolution is required. The estimated $\delta r$ is comparable to the pixel size in the movies (57 nm in Movie M1 and 43 nm in M3,4). We thus conclude that the spatial resolution $\delta r$ of the attained $\theta(r)$ maps is of the order of 4 to 5 moiré supercells (13 nm each).

**SI7. Local quantum Hall measurement in device A**

Figure S7 presents the local $B_z^{ac}$ measurement, with SOT parked at a fixed position, along with the global transport $R_{xx}$ measurement in device A at $B_a = 1.19$ T. Alternating compressible and incompressible states in the region under the tip lead to a series of peaks in $B_z^{ac}$, with sharp peaks corresponding to incompressible strips carrying $I^{inc}$. The sign of the incompressible peaks is determined by the sign of $\sigma_{yx}$, with $B_z^{ac} > 0$ ($< 0$) for electron (hole) doping. In Figs. 1f and 2a the $B_z^{ac}$ signal for p doping was multiplied by minus one for clarity. The spacing between adjacent peaks reflects the degeneracy of the LL. The dispersive band (shaded yellow), exhibits a sequence of 4-fold and 8-fold degeneracies. In the flat band we find 4-fold degenerate levels around $n_e = 0$, 2-fold degeneracy near $n_e = \pm n_s/2$, and 1-fold degenerate levels near $n_e = -3n_s/4$ (see Fig. 1f for 1-fold degenerate levels near $n_e = +3n_s/4$). Evaluation of the local $n_s$ allows the extraction of the local twist angle, $\theta = 1.136 \pm 0.005°$ as described in SI6. In contrast to the sharp $B_z^{ac}$ local peaks, oscillations in $R_{xx}$ are hardly visible due to $\theta(r)$ disorder and the fact that the MA regions in device A do not extend over the entire device area.

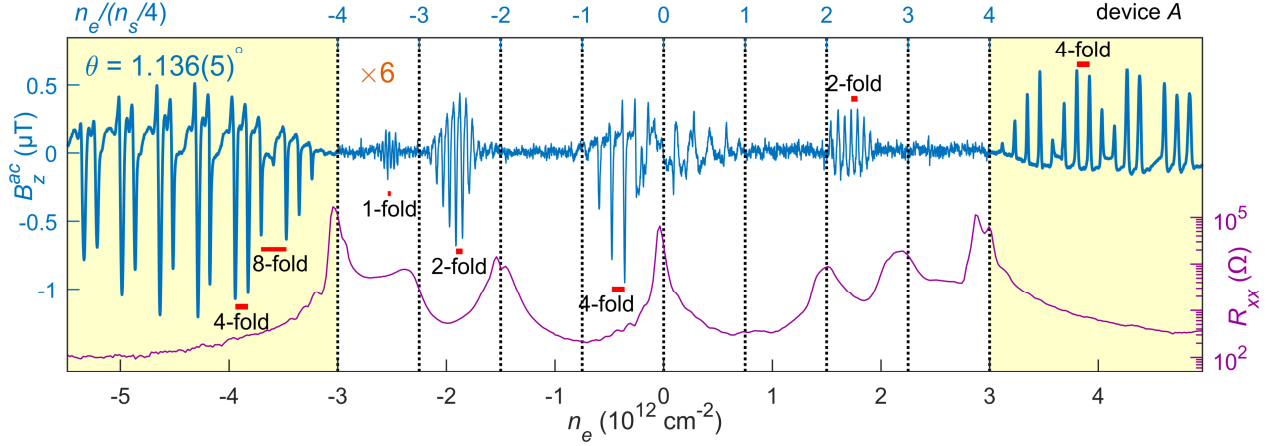

**Fig. S7. Resolving the local quantum Hall states in flat and dispersive bands in device A.** Global $R_{xx}$ (right axis) and local $B_z^{ac}$ (left) measured at a point in the bulk of device A vs. electron density $n_e$ at $B_a = 1.19$ T. The sharp $B_z^{ac}$ peaks reflect $I^{inc}$ current in incompressible strips with sign determined by the sign of $\sigma_{yx}$, magnitude by the LL energy gap, and separation by the LL degeneracy (red bars). The dispersive bands are shaded in yellow and the signal in flat bands amplified 6 times for clarity.

### SI8. Movies M1-M5

**Movie M1.** Zoom on the central part of the Hall bar in device A (purple rectangle in Fig. 3a) upon sweeping $V_{bg} = -8.58$ to $-11.50$ V in fine steps of $\Delta V_{bg} = 40$ mV through the bottom of the $p$ dispersive band. No LLs are visible in the first few frames when $|n_e| < n_s(r)$ everywhere. On increasing $|V_{bg}|$, the hole concentration reaches $n_s(r)$ first where $\theta(r)$ is the lowest, which occurs in the lower-right corner (the darkest region in Fig. 3b and the 3D triangular structure in Fig. 3d). There the incompressible strips (light blue to red-yellow) are formed with arc-like shape and "climb" the amphitheater-like $\theta(r)$ landscape following the equi-$\theta(r)$ contours. Note that the incompressible arcs do not form closed contours as expected in conventional QH state and the equilibrium $I^{inc}$ currents flowing along the arcs apparently "return" through the counterflowing compressible $I^{com}$ (dark blue). Similar dynamics occurs in the top-right corner (dark brown in Fig. 3b) where parallel stripes appear and climb towards bottom-left until they reach a saddle point. There stripes approaching from opposite slopes merge at the saddle point and then split and climb the other two slopes with perpendicular orientation (bright yellow in Fig. 3b), as seen in frames $V_{bg} = -10.46$ V to $-10.7$ V and then repeated for the next LL at $V_{bg} = -11.22$ V to $-11.50$ V.

**Movie M2.** The movie presents a sequence of large area $B_z^{ac}$ images in the $p$ dispersive band in device A upon sweeping $V_{bg} = -16.4$ to $-17.0$ V in coarse steps of $\Delta V_{bg} = 0.1$ V (see SI13 for other parameters). In this range, $\varepsilon_F$ lies relatively high in the $p$ dispersive band corresponding to $|n_e| \approx 1.7 n_s$ (see Fig. 2a). The edges of the device are indicated by the black lines and the first frame of the movie is described in Fig. 3a. The movie shows the evolution of the compressible (light blue to red-yellow) and incompressible (dark blue) strips upon increasing the hole concentration. Note that large parts of the sample, including the entire top part, bottom left area, and smaller regions in the center containing bubbles, do not show LLs and MATBG physics at all. These are the areas that are highly disordered or have a very different twist angle.

**Movies M3,4.** The two movies show the evolution of the LLs in the *p* and *n* dispersive bands in device *B* (see SI13 for parameters). The data in Movie M3 is negated so that the incompressible strips appear bright in all the movies. Similarly to device *A* the LLs climb the $\theta(r)$ slopes which are particularly large in the top and bottom regions of device *B* (dark brown in Fig. 3f). The steep slopes of LLs in these regions are clearly visible in the tomography (Fig. S8c).

**Movie M5**. The movie shows an alternate method for visualizing the data presented in Movie M1, as described in SI9. The $B_z^{ac}$ data is shown in the $y - V_{bg}$ plane at different $x$, then in $x - V_{bg}$ plane at different $y$ and finally in the $x - y$ plane at different $V_{bg}$. The angle inhomogeneity is seen as a change in $V_{bg}$ of each incompressible stripe (red) in the $x - V_{bg}$ or $y - V_{bg}$ planes. Regions which do not have incompressible regions for any value of $V_{bg}$ correspond to bubbles seen in the AFM image (Figure 3a inset).

**SI9. Landau level tomography and twist angle mapping**

In order to map the local twist angle, a series of $B_z^{ac}(r)$ area scans were performed upon varying $V_{bg}$. This results in a 3D dataset with two spatial dimensions and one $V_{bg}$ (or equivalently $n_e$) axis. Each LL energy gap forms a 2D manifold in this 3D space with a peak in $B_z^{ac}$ signal (bright in Fig. S8). The manifolds of the lowest LLs in the dispersive bands trace the manifold of the bottom of the dispersive band, $n_s(r)$, and are displaced vertically from it by the degeneracy of the LLs, thus providing the means for mapping the local $n_s(r)$ and hence the local $\theta(r) = a\sqrt{\sqrt{3}n_s(r)/8}$. The 3D space was mapped with pixel size of ~50 nm and $V_{bg}$ spacing between successive scans, $\Delta V_{bg} \cong 40$ mV, which allows mapping $\theta(r)$ with accuracy $\delta\theta = \pm 0.001°$ (see SI6).

For device *A*, the tomographic imaging was acquired for the *p* dispersive band for $V_{bg}$ spanning $-8.58$ V to $-11.50$ V with $\Delta V_{bg} = 40$ mV (Movie M1). The spacing between adjacent 4-fold levels at $B_a = 1.22$ T was $0.39$ V $\cong 10\Delta V_{bg}$. In this device, the spatial variation of the charge neutrality voltage $V_{bg}^{CNP}(r)$ was found to be very small (Fig. 2a) and therefore $n_s(r)$ was derived from the 3D data assuming a constant $V_{bg}^{CNP}$. Representative slices of the 3D dataset are shown in Figs. S8a,b. At $V_{bg} = -8.5$ V, the Fermi level resides in the flat band for all points in space, and at $V_{bg} = -11.5$ V, $\varepsilon_F$ is in the dispersive band. As $\varepsilon_F$ moves through the bottom of the dispersive band, it crosses four 4-fold degenerate LLs above $n_s$ followed by an 8-fold degenerate LL. The black line in Fig. S8a traces the $N = -4$ incompressible $I^{inc}$ peak revealing $\theta(r)$ gradients with occasional small jumps in the twist angle. Note that at the jump positions, the intensity of the $B_z^{ac}$ signal is suppressed due to pinning of the LLs at the $\theta(r)$ steps, which reduces the amplitude of the spatial *ac* displacement $x_0^{ac}$ and hence the intensity of $B_z^{ac}$ (see SI4).

Device *B* exhibited stronger charge inhomogeneity and hence the 3D tomographic imaging was acquired for both *p* and *n* dispersive bands (Movies M3 and M4) and $n_s(r)$ was derived from the separation between the corresponding LLs in the two bands as described schematically in Fig. 1c. The tomographic data of both samples is available on [30].

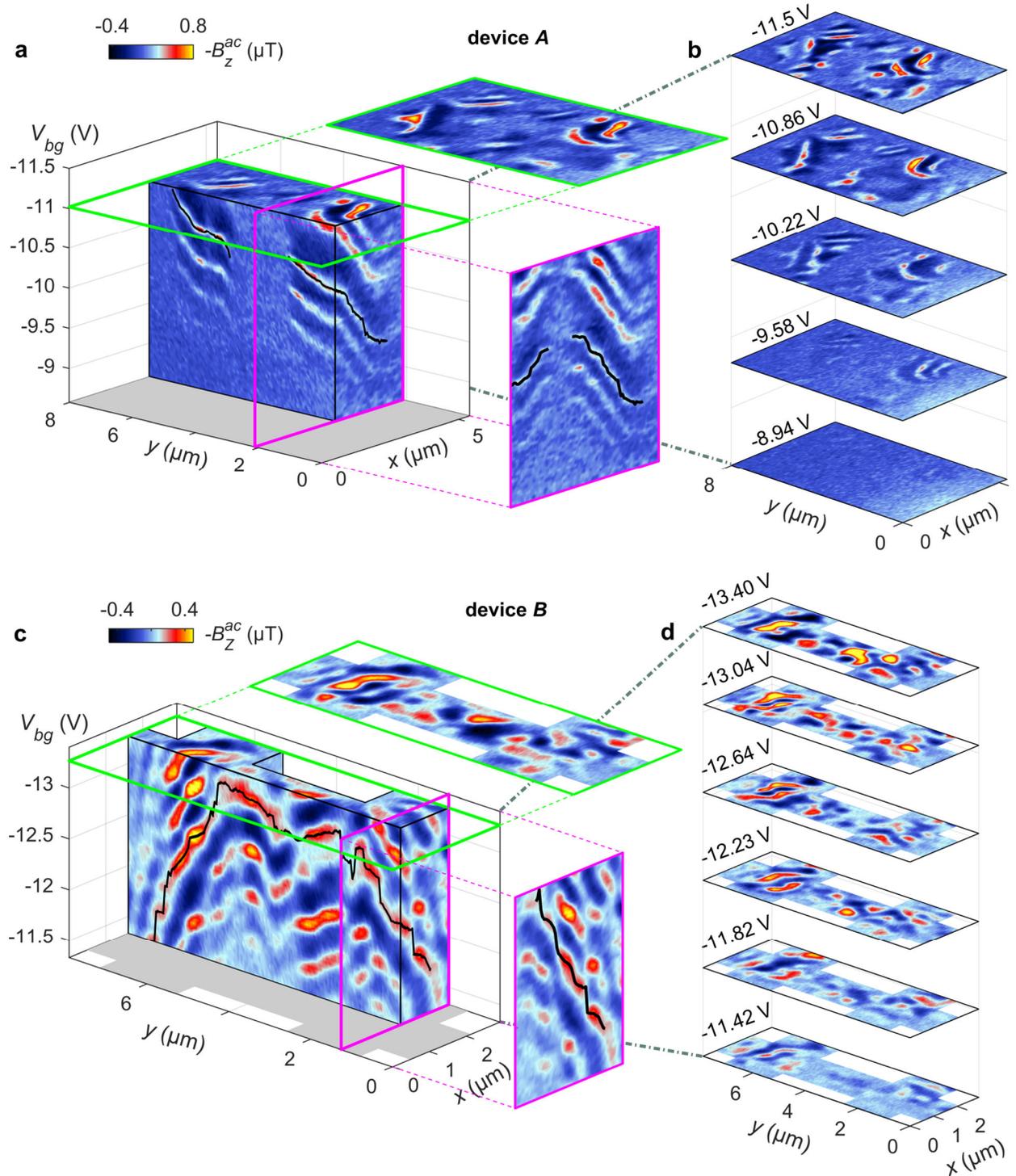

**Fig. S8. Landau level tomography.** (a) Slices of the 3D dataset $B_z^{ac}(x, y, V_{bg})$ along various planes for device A. The bright signals denote the 2D manifolds tracing the incompressible states. The black lines trace the $N = -4$ incompressible manifold used to determine $n_s(x,y)$ and $\theta(x,y)$. It separates 4-fold degenerate LLs below it from an 8-fold degenerate LL above it (wide dark blue band). The region in the center of the sample showing no LLs corresponds to the grey-blue area in Fig. 3b where no MATBG physics is resolved. (b) Representative horizontal slices of the data from Movie M1 showing the evolution of the LLs with $V_{bg}$. (c) Same as (a) for device B. For the range of gate voltages shown, $\epsilon_F$ lies in the $p$ dispersive band for the entire sample. The black lines show an example of trace of the incompressible manifold lying above an 8-fold degenerate LL. (d) Representative horizontal slices of the data from Movie M3. Interactive interface for tomographic visualization of the data is available on [30].

## SI10. Mapping of the charge disorder

Similarly to the mapping of the twist angle disorder through $n_s(r) = C(V_{ns}(r) - V_{-ns}(r))/2$, the tomographic imaging also allows mapping of the charge disorder $\delta n_d(r) = C(V_{ns}(r) + V_{-ns}(r))/2 - \bar{n}_d$, as presented in Fig. 3h. Figure S9 shows the histogram of $\delta n_d(r)$ in device B along with a Gaussian fit exhibiting standard deviation $\Delta n_d = 2.59 \times 10^{10}$ cm$^{-2}$.

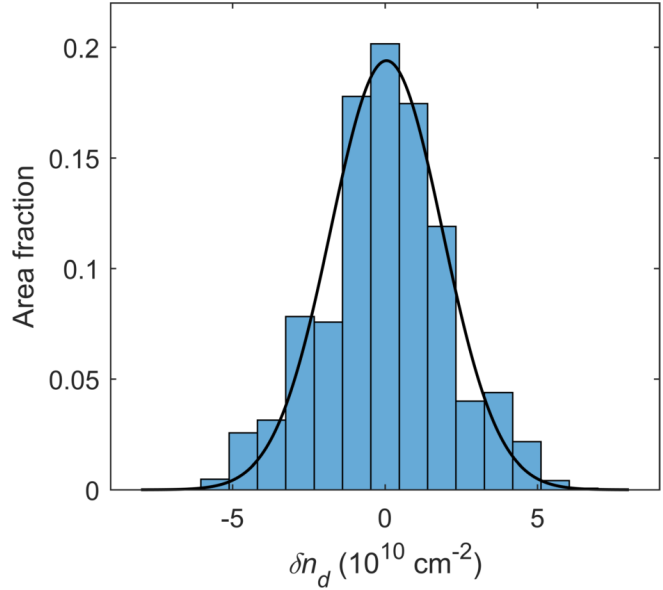

**Fig. S9. Histogram of the charge disorder in device B.** Histogram of $\delta n_d(r)$ data from Fig. 3h along with a Gaussian fit (black) with standard deviation of $\Delta n_d = 2.59 \times 10^{10}$ cm$^{-2}$.

Note that the tomographic method allows mapping of the twist angle and charge disorders only in the MA regions where LLs are present. In device A, a significant part of the sample did not show MA physics (Movies M1,M2 and Figs. 3a,b) while the regions of MA revealed very low charge disorder with an estimate standard deviation $\Delta n_d \approx 1.3 \times 10^{10}$ cm$^{-2}$ as attained by several 1D scans like e.g. Figs. 2a,d. We therefore performed tomographic imaging of only the $p$ dispersive band which does not allow extracting the full 2D map of $\delta n_d(r)$ in device A. Neglecting this low level of charge disorder introduces an error in the derived $\theta(r)$ map of device A of $\delta\theta \lesssim 0.0015°$, which is negligible compared to the span of $\theta(r)$ in Fig. 3b.

## SI11. Band structure calculations and Landau level crossings

The band structure of twisted bilayer graphene can be computed from an effective continuum Hamiltonian, which reads [9,10,46–48]

$$H^{(\xi)} = \begin{pmatrix} H_1^{(\xi)} & U^\dagger \\ U & H_2^{(\xi)} \end{pmatrix},$$

where $H_i^{(\xi)}$ is the valley dependent monolayer graphene Hamiltonian for layer $i$,

$$H_i^{(\xi)} = -\hbar v_F \left(k - K_i^{(\xi)}\right) \cdot (\xi\sigma_x, \sigma_y),$$

with Fermi velocity $v_F$ and $\xi = \pm 1$, indicating the positive and negative valleys, $K_i^{(\xi)}$ is the k-space location of the respective Dirac points in layer $i$, and $U$ is the interlayer coupling, which reads [48–51]

$$U = \begin{pmatrix} u & u' \\ u' & u \end{pmatrix} + \begin{pmatrix} u & u'\omega^* \\ u'\omega & u \end{pmatrix} e^{i\xi G_1^M \cdot r} + \begin{pmatrix} u & u'\omega \\ u'\omega^* & u \end{pmatrix} e^{i\xi (G_1^M + G_2^M) \cdot r}.$$

Here, $u = 0.0797$ eV and $u' = 0.0975$ eV [49] are coupling constants that give the strength of the interaction between like $(A \leftrightarrow A, B \leftrightarrow B)$ and opposing $(A \leftrightarrow B)$ sublattices in the two layers, the difference of which accounts for out-of-plane corrugation, and $\omega = e^{2\pi i/3}$. The Moiré reciprocal lattice vectors, $G_j^M = a_j^{(1)} - a_j^{(2)}$, are given by the difference between the reciprocal lattice vectors in the upper ($a_j^{(1)}$) and lower ($a_j^{(2)}$) layers.

Magnetic field effects can be included by making the substitution $k \to k + eA/\hbar$ in the effective Hamiltonian. Here, $A$ is the vector potential which is related to the static magnetic field via $B = \nabla \times A$. In general, the band structure in a magnetic field cannot be computed because the addition of a spatially dependent vector potential breaks translational invariance. However, at certain values of the magnetic field – specifically when $(SB/h)/e = p/q$, where $p$ and $q$ are co-prime integers and $S$ is the area of the unit cell – a "magnetic" unit cell can be introduced whereupon it becomes possible to solve the Schrödinger equation using the corresponding "magnetic" Bloch conditions [50]. One is then able to construct a Hamiltonian matrix in the basis of the monolayer graphene Landau levels [51,52]. Although the Landau levels basis is unbounded, one can truncate the Hamiltonian matrix at an energy where the higher energy LLs only weakly affect the low energy spectrum. This cut-off energy must be significantly larger than the interlayer coupling characterized by the coupling constants $u$ and $u'$. The resulting finite matrix can then be diagonalized. This results in a band structure diagram in terms of $p/q$, which is directly related to the strength of the magnetic field and indirectly related to the twist angle as the Moiré unit cell area, $S = \sqrt{3}a^2/(8\sin^2(\theta/2))$, is proportional to the twist angle $\theta$. For varying magnetic field or twist angle the bands are computed for each individual parameter value assuming that these values are homogeneous throughout the material.

Level crossings in the band structure are observed as one varies the magnetic field or the twist angle. This is owing to the "Rashba-like" splitting of the dispersive bands. In general, this type of splitting leads to two Landau level series, largely overlapping in energy, which cross as a function of magnetic field (Fig. S10a) [53]. Similarly, LL crossings are also observed as a function of $\theta$ (Fig. S10b) as is the case in the experimental data. This is owing to the evolution of the "Rashba-like" splitting with $\theta$ (Figs. S10c-e).

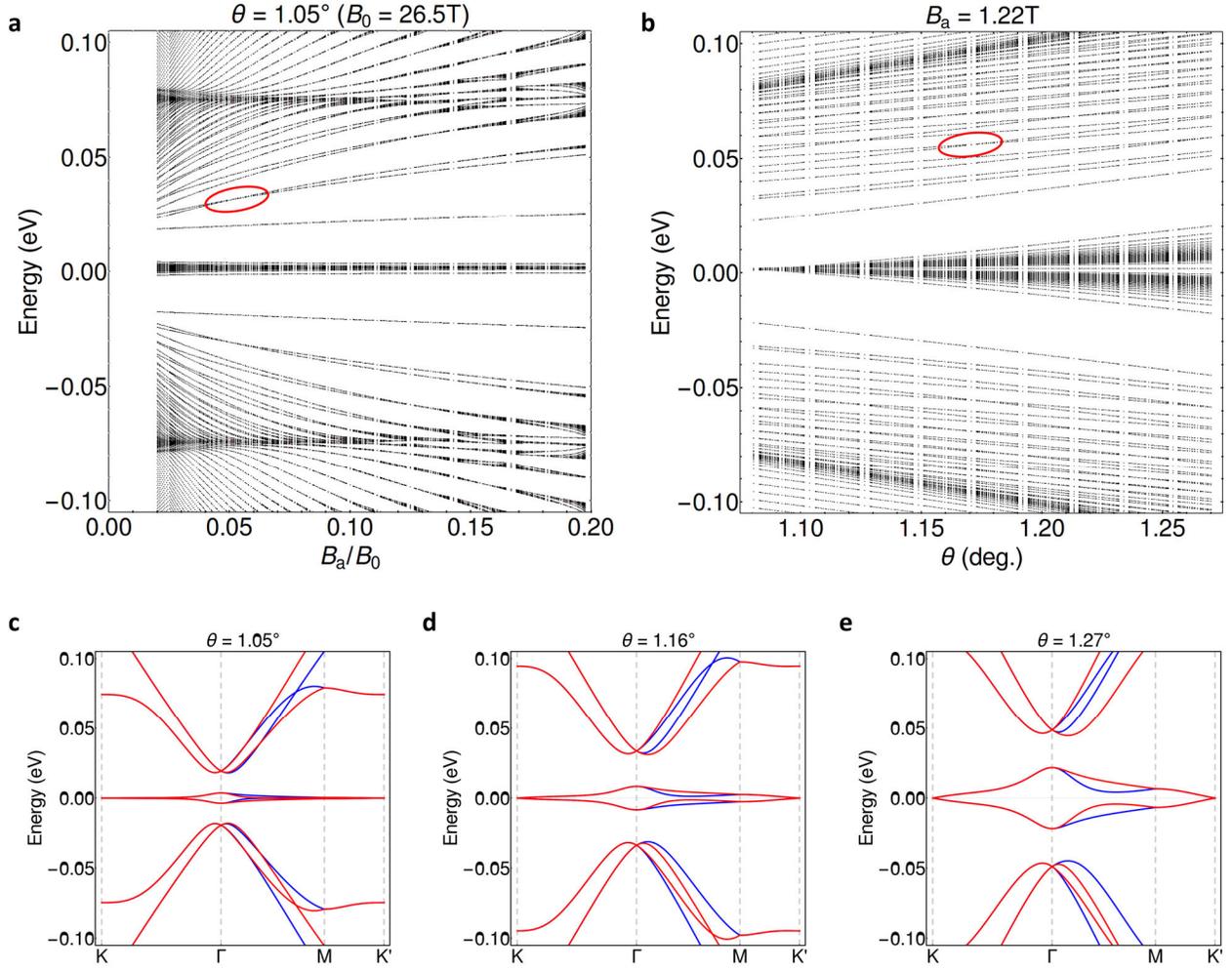

**Fig. S10. Dispersive bands Landau level crossings.** (**a**) Numerically calculated LL energies as a function of magnetic field for a fixed $\theta = 1.05°$. An example of level crossing is highlighted in red. (**b**) Numerically calculated LL energies as a function of $\theta$ for a fixed $B_a = 1.22$ T. An example of a level crossing is highlighted in red. (**c-e**) The $B_a = 0$ band structure of bilayer graphene for $\theta = 1.05°$ (**c**), 1.16° (**c**), and 1.27° (**c**). The blue and red lines indicate the bands that arise from the positive and negative valleys, respectively.

### SI12. Numerical electrostatic simulations

For results presented in Fig. 4, COMSOL simulations were used for solving electrostatic equations for the potential $V$ and charge density $\rho = -en_e$ at $B_a = 0$ and in the QH state at $B_a = 1.22$ T. The simulations included a backgate at a constant electric potential $V_{bg}$ and a grounded MATBG in a 3×0.5 μm² $x$-$z$ box, assuming translation invariance along the $y$ axis, with boundary conditions of $E_\perp = 0$ on the box's external surfaces. An iterative self-consistent solution for $V(x, y, z)$ and $\rho$ was obtained, satisfying the following conditions: (1) The electric potential $V(x, y, z)$ depends on $\rho$ through $\nabla \cdot \boldsymbol{E} = \rho/\varepsilon_r \varepsilon_0$ and $\boldsymbol{E} = -\nabla V$ with given $V_{bg}$, where $\varepsilon_r$ is the relative permittivity (we took $\varepsilon_r = 4$ for hBN) and $\varepsilon_0$ is the vacuum permittivity. (2) $\rho$ depends on $V$ through the integrated density of states $\mathcal{N}_e(\mu; \theta)$, where $\mu(x, y) = -qV(x, y)$ and $q = \pm e$ is the carrier charge (negative sign for $\mu > 0$). The integrated density of states $\mathcal{N}_e(\mu; \theta)$ was calculated for $B = 0$ and $B \neq 0$ as described in SI11.

Once $V(x,y,z=0)$ and $\rho(x,y)$ were found in the plane of the MATBG, the incompressible surface currents were calculated using $\mathbf{J}^{inc} = -\sigma \nabla V$, where $\sigma_{xy}(x,y) = -\sigma_{yx}(x,y) = -\nu(x,y)e^2/h$ and $\sigma_{xx} = \sigma_{yy} = 0$ are the components of the conductivity tensor $\sigma$.

**SI13. Measurement parameters**

All the measurements were carried out at $T = 300$ mK in out-of-plane applied magnetic field $B_a$.

Fig. 1c, S6a and S7: Device A, $B_a = 1.19$ T, SOT diameter 220 nm, scan height 100 nm, $V_{bg}^{ac} = 15$ mV rms, $\Delta V_{bg} = 4.7$ mV, acquisition time 6 s per point, total acquisition time 12 hours.

Fig. 1f: Device B, $B_a = 1.08$ T, SOT diameter 250 nm, scan height 40 nm, $V_{bg}^{ac} = 20$ mV rms, $\Delta V_{bg} = 6.25$ mV, acquisition time 6 s per point, total acquisition time 8 hours.

Fig. 2: Device A, $B_a = 1.22$ T, SOT diameter 220 nm, scan height 60 nm, $V_{bg}^{ac} = 35$ mV rms, pixel size 26 nm, 160 ms per pixel, total acquisition time 21.4 hours.

Fig. 3a and movie M2: Device A, $B_a = 1.16$ T, SOT diameter 220 nm, scan height 110 nm, $V_{bg}^{ac} = 80$ mV rms, pixel size 60 nm, 60 ms per pixel, acquisition time 60 minutes per frame.

Movie M1 and Figs. 1d, S6b and S8a,b: Device A, $B_a = 1.22$ T, SOT diameter 220 nm, scan height 60 nm, $V_{bg}^{ac} = 35$ mV rms, pixel size 57 nm, 60 ms per pixel, acquisition time 30 minutes per frame.

Fig. 3e: Device B, $B_a = 1.08$ T, SOT diameter 250 nm, scan height 140 nm, $V_{bg}^{ac} = 60$ mV rms, pixel size 50 nm, 60 ms per pixel, acquisition time 33 minutes.

Movies M3-4 and Fig. S8c,d: Device B, $B_a = 1.08$ T, SOT diameter 250 nm, scan height 70 nm (M3) and 80 nm (M4), $V_{bg}^{ac} = 60$ mV rms, pixel size 43 nm, 60 ms per pixel, acquisition time 25 minutes per frame.

Fig. S2a: Device A, $I_{ac} = 10$ nA, $V_{bg} = -17$ V to 17 V. Silicon backgate 50 V.

Figs. 1d and S2b: Device B, $I_{ac} = 10$ nA, $V_{bg} = -15$ V to 15 V.

Fig S3a: Device A, $I_{ac} = 5$ nA, $V_{bg} = -6$ V to $-3.5$ V. Silicon backgate 50 V.

Fig S3b: Device B, $I_{ac} = 4$ nA, $V_{bg} = -7$ V to $-4$ V.

Fig S3c: Device B, $I_{ac} = 10$ nA, $V_{bg} = -6.5$ V to $-4.7$ V, $B_a = 0$ T.